\documentclass[preprint]{imsart}

\usepackage[T1]{fontenc}
\usepackage[utf8]{inputenc}
\usepackage[english]{babel}
\usepackage{amsfonts}
\expandafter\let\csname equation*\endcsname\relax
\expandafter\let\csname endequation*\endcsname\relax
\usepackage{amsmath}
\usepackage{amsthm}
\usepackage{amssymb,amsxtra,amscd}
\usepackage{mathrsfs,ae,dsfont,algpseudocode,algorithm}
\usepackage{graphicx}
\usepackage[normalem]{ulem}
\usepackage{nicefrac,ucs,xcolor,cmll}
\usepackage{natbib}
\usepackage{layout}
\usepackage{fancybox,framed}
\usepackage{subfig}
\usepackage{slashbox,hhline}
\usepackage{algorithm}

\newif\ifarxiv
\newif\ifaoas

\renewcommand\fbox[1]{\Ovalbox{#1}}

\def\plus#1#2{\vrule height#1pt width0pt depth#2pt}
\def\@{\hskip.8pt}
\def\?{\hskip.3pt}

\newcommand{\abs}[1] {|#1|}

\newcommand{\de}{\partial}
\renewcommand{\leq}{\leqslant}
\renewcommand{\geq}{\geqslant}

\DeclareMathAlphabet{\mathpzc}{OT1}{pzc}{m}{it}

\def\b{\mathbf{b}}

\def\C{\mathds C\/}

\newcommand{\D}[1]{\frac{D #1}{D t}}

\def\h#1{\hat{#1}}

\def\J{\mathcal{J}}

\def\L{\mathfrak{L}\/}

\def\N{\mathbb{N}\/}
\def\narc#1{\text{\@ \tiny $(#1)$}}

\def\nc{N_C}
\def\nd{N_D}
\def\ni{N_I}
\def\np{N_P}
\def\ns{N_S}
\def\Nc{\text{\tiny $\nc$}}
\def\Nd{\text{\tiny $\nd$}}
\def\Ni{\text{\tiny $\ni$}}
\def\Np{\text{\tiny $\np$}}
\def\Ns{\text{\tiny $\ns$}}
\def\nutilde{\?{\scriptstyle\mbox{\tiny\raise-1.6ex\hbox to1pt{$\sim\hss$}}\nu\@\?}}
\def\Nutilde{\raise-1ex\hbox to1pt{$\scriptstyle\sim\hss$} \nu}

\def\P_#1{\mathcal P_{#1}\/}

\def\r{\mathbf{r}}
\def\R{\mathds{R}\/}

\def\X{\mathcal X\/}

\def\d#1/d#2{\frac{d\/#1}{d\/#2}}
\def\de#1/de#2{\frac{\partial\/#1}{\partial\/#2}}
\def\SD#1/de#2/de#3{\ifx#2 \frac{\plus02\partial^{\@\@2}#1}
    {\plus90\partial\@#3^{\@2}} \else\frac{\plus02\partial^{\@\@2}#1}
    {\partial\?#2\partial\?#3}\fi}
\def\TD#1/de#2/de#3/de#4{\ifx#2 \frac{\plus02\partial^{\@\@3}#1}%
    {\partial\?{#3}^2\partial\?#4} \else\frac{\plus02\partial^{\@\@3}#1}%
    {\partial\?#2\partial\?#3\partial\?#4}\fi}
\def\D#1/D#2{\frac{D\/#1}{D\/#2}}
\def\oD#1/d#2{\textstyle{\text{\large$\d{#1}/d{#2}$}}}
\def\OD#1/d#2{\textstyle{\text{\Large$\d{#1}/d{#2}$}}}
\def\De#1/de#2{\textstyle{\text{\Large$\de{#1}/de{#2}$}}}
\def\sd#1/de#2/de#3{\textstyle{\text{\large$\SD{#1}/de{#2}/de{#3}$}}}
\def\sD#1/de#2/de#3{\textstyle{\text{\Large$\SD{#1}/de{#2}/de{#3}$}}}
\def\sOD#1/d#2{\textstyle{\text{\large$\d\@^2{#1}/d{#2\?^2}$}}}

\def\dd#1/dd#2{\frac{\delta\/#1}{\delta\/#2}}

\theoremstyle{plain}

\newtheoremstyle{note}%
{15pt}%
{3pt}%
{\small}%
{}%
{\bfseries}%
{:}%
{.5em}%
{}%

\theoremstyle{note}

\numberwithin{equation}{section}
\theoremstyle{plain}

\catcode`\@=11 \@addtoreset{equation}{section}
\def\theequation{\thesection.\arabic{equation}}
\catcode`\@=12

\catcode`\@=11
 \catcode`\@=12

\def\Ref#1#2{\if#2)\ref{#1}#2\else\ref{#1}\@#2\fi}

\def\inizio#1{\ifx)#1)\endzag\else\hskip.1pt,\let\zag=\Zag\fi\zag}
\def\zig#1{\ifx)#1)\endzag\else\hskip.1pt,\fi\zag}
\def\zag{\relax}
\def\Zag#1{\@#1\zig}
\def\endzag{\let\zag=\relax}

\def\eref#1#2{\ref{#1}\ifx,#2\hspace{.1pt},\let\inizio=\Zag\let\zag=\Zag\else\@#2\fi\inizio}

\usepackage{fullpage}
\begin{document}

\begin{frontmatter}

\title{Bayesian Multi--Dipole Modeling of a Single Topography in MEG by Adaptive Sequential Monte Carlo Samplers}
\runtitle{Bayesian Multi--Dipole Modeling of a Single Topography in MEG by ASMC}
\begin{aug}
\author{\fnms{Alberto} \snm{Sorrentino}\thanksref{t1}\ead[label=e1]{sorrentino\@@\@dima.unige.it}},
\author{\fnms{Gianvittorio} \snm{Luria}\ead[label=e2]{luria\@@\@dima.unige.it}}, \\
\and \author{\fnms{Riccardo} \snm{Aramini}\ead[label=e3]{aramini\@@\@dima.unige.it}}\\
\thankstext{t1}{A.S. was partially supported by a Marie Curie Intra European Fellowship
within the 7th European Community Framework Programme.}

\runauthor{A. Sorrentino et al.}
\affiliation{Universit\`a di Genova}

\address{Dipartimento di Matematica -- Universit\`a di Genova\\
Via Dodecaneso, $35$ -- $16146$ Genova (\@Italia\@)\\
\printead{e1}\\
\printead{e2}\\
\printead{e3}\\
\phantom{E-mail:\ }}
\end{aug}

\begin{abstract}
In the present paper, we develop a novel Bayesian approach to the problem of estimating neural currents in the brain
from a fixed distribution of magnetic field (called \emph{topography}), measured by magnetoencephalography.
Differently from recent studies that describe inversion techniques, such as spatio-temporal regularization/filtering, 
in which neural dynamics always plays a role, we face here a purely static inverse problem. 
Neural currents are modelled as an unknown number of current dipoles, whose state space is described in terms of a variable--dimension model.
Within the resulting Bayesian framework, we set up a sequential Monte Carlo sampler to explore the posterior distribution.
An adaptation technique is employed in order to effectively balance the computational cost and the quality of the sample approximation.
Then, both the number and the parameters of the unknown current dipoles are simultaneously estimated.
The performance of the method is assessed by means of synthetic data, generated
by source configurations containing up to four dipoles. Eventually, we describe the results
obtained by analyzing data from a real experiment, involving somatosensory evoked fields,
and compare them to those provided by three other methods.
\end{abstract}

\end{frontmatter}

\section{Introduction}
Magnetoencephalography (\/M\/E\/G\/) is a non--invasive functional
neuroimaging technique \citep{haetal93} that records
the weak magnetic fields produced by neural currents in the brain. A modern M\/E\/G device,
made of a helmet--shaped array of a few hundred of SQUID sensors, has two major advantages\/:
it measures the most direct consequence of the brain electrical activity,
and it does so with the outstanding temporal resolution of the order of
the millisecond, only matched by its closest akin
electroencephalography (\/E\/E\/G\/).
By contrast, the
more widely used functional Magnetic Resonance Imaging (\@fM\/R\/I\@)
provides measures of the by--product of complex metabolic mechanisms at a rate of a second.
The high temporal resolution is a crucial feature of M\/E\/G, since it enables to investigate the neural dynamics
in a wide variety of conditions, both normal and pathological, where
neural oscillations are thought to play a relevant role. Typical
examples are Alzheimer's disease \citep{stetal09}, Parkinson's disease \citep{stetal07} and epilepsy \citep{udetal12}.
Moreover, compared to E\/E\/G, M\/E\/G is less influenced by the inhomogeneities and anisotropies inside
the head, which have a high inter--individual variability and can hardly be taken into account within the models.

From a mathematical viewpoint, the estimation of neural currents from M\/E\/G data is known
to be an ill--posed inverse problem \citep{sa87}, suffering from non--uniqueness of the
solution \citep{dafoka05, fokuma04}. 
%
Source estimation from M\/E\/G data is therefore a challenging, yet
worth investigating, problem.

Generally speaking, the above inverse problem may be addressed by two different
approaches: the \emph{dipolar} and the \emph{distributed} source model \cite{haetal93}.
In the latter framework, neural currents are modelled as continuous vector fields,
discretized on a dense mesh, and the data depend linearly on the unknowns.
Among the regularization techniques used,
$\@L^1$--penalty terms are becoming increasingly popular \citep{chetal12}, as
they provide sparser estimates that more closely  reflect the
neurophysiological evidence of the neural generators being rather
focal. By contrast, this fact is naturally coded in dipolar source models, whereby
neural currents are modelled as the superposition of a small
number of point--like sources, called \emph{current dipoles}.
However, estimation of dipole parameters from M\/E\/G data is more
difficult, since the relationship between the source positions and
the generated magnetic field is non--linear. Traditionally,
``\/fitting\/'' algorithms are used \citep{sa10} to produce maximum
likelihood estimates of individual dipoles, but these methods need
careful initialization and may get stuck in local minima. Recently,
spatio--temporal analysis of M\/E\/G data has been performed by means of
Bayesian Monte Carlo techniques, such as particle filters (\/PF\/) \citep{sovoka03,soetal09} and
Markov Chain Monte Carlo (\/MCMC\/) \citep{juetal05}. Given a time--series of MEG data,
these methods approximate the posterior distribution for a time--varying set of current dipoles by exploiting the expected temporal
continuity of the underlying generators. They have been shown to perform reasonably well with evoked responses, although PF
exhibit relatively high variance at the appearance of a new source, since the importance distribution is too different
from the posterior: this typically leads to slightly inaccurate localization of the onset, and a localization error decreasing
quickly in the very first time points, while particles concentrate in the high--probability region \cite{so10,soetal13}.

In this paper, we take an apparent step back, and develop a new Bayesian approach to solve
the seemingly simpler, static inverse problem of determining
the number of dipoles and their parameters from a single spatial distribution of the magnetic field (\/\emph{topography}\/).
Importantly, such single topography may be obtained not only as a single time point of an MEG time--series, but also,
for instance, by picking the magnetic field at a given frequency after Fourier transform, or a
single spatial component after Independent Component Analysis (\/ICA\/); these data--processing
techniques can be used, even in combination \citep{hyetal10}, for analyzing resting--state \citep{bretal11} and event--related
\citep{suta06} data.

The proposed method extends the applicability of Bayesian multi--dipole models to a variety of conditions
where the only inversion methods currently available are linear inversion or the  above mentioned dipole fit.
It ought to be stressed that dipole estimation from a single topography can not be successfully performed by means
of PF, not even the one designed to estimate static dipoles, described in \cite{soetal13}.
Indeed, PF applied to a single topography would result in a single importance sampling step, with unsatisfactory results
due to the difference between the prior and the posterior distribution. A seemingly reasonable option would be to re-iterate a
PF with the same data (i.e., simulating a time--series which is constantly identical to the single topography).
However, this would be rather questionable from a Bayesian point of view: from the second iteration onwards the posterior distribution
of the previous iteration, convolved with the transition kernel, would be regarded as the prior for the current iteration, without being 
such in a Bayesian sense.
In fact, while the absence of dynamics may breathe simplicity, the problem addressed in this paper and that solved by PF
are comparably hard: although the state space is now smaller, the data are also fewer, and the number of sources has to
be estimated at once, while in a typical M\/E\/G time--series dipoles will mostly appear one at a time.

Within our framework, the state space of current dipoles is described by making use of a variable--dimension model \cite{camory05, roca04}.
We choose uninformative prior distributions for all parameters but the number of sources, which is assigned a Poisson distribution;
this choice grounds on the consideration that dipole models implicitly assume a small number of sources.
On the contrary, the likelihood is typically highly informative and naturally tends to high--dimensional configurations.
The resulting posterior is therefore a complex,
possibly multi--modal density in
a high--dimensional space, and negligible in most of the state space\/: it is therefore not analytically
tractable, nor easy to sample.

To explore a posterior distribution with such a complex structure, 
 we apply a recent class of algorithms, called \emph{Sequential Monte Carlo}
(\/SMC\/) \emph{samplers} \citep{dedoja06}.
The basic idea on which they are grounded is that of building a
sequence of artificial distributions
$\{\pi_i\}_{i\@=1}^{\!\Ni}\,$ that smoothly moves from a tractable
$\pi_1$ to the target $\pi \equiv \pi_{\Ni}\@$. By sampling $\pi_1$
and letting the samples evolve gradually to account for the
difference between two subsequent distributions, one eventually
obtains a collection of points (\/\emph{particles}\/) that is
actually distributed according to $\pi$.

The choice of the sequence is clearly a key--point  in the
definition of an SMC sampler; in this paper, we describe an adaptive
approach, called \emph{Adaptive Sequential Monte Carlo} (\/ASMC\/) \emph{sampler},
whereby the sequence is determined on--line, with an empirical measure of
the distance between two subsequent distributions. In this way, we
optimize the trade--off between the computational speed and the
quality of the approximation.

We apply the resulting algorithm to a number of synthetic MEG
topographies, constructed to assess its performances under a range
of experimental conditions.
We also show that the method performs reliably on real data, by making
use of single time points selected from the
responses to somatosensory stimulation, which is a well understood case.\\

The paper is organized as follows. In Section \ref{S_2} we provide the
mathematical formulation of the MEG problem and its statistical model.
To improve readability, the detailed definition of the state--space
is given in \ref{S_A_1}.
Section \ref{S_3} recalls the basic concepts of generic SMC samplers and describes
their specific implementation for MEG, as well as the corresponding
adaptation technique. In Section \ref{S_4} an extensive analysis of a large
number of synthetic data sets is performed. Section \ref{S_5} is concerned
with the application of the same algorithm to a real MEG data set.
Finally, we propose a discussion of the results and our conclusions in
Section \ref{S_6}.

\section{The MEG inverse problem}\label{S_2}
\subsection{Mathematical model of M\@E\@G}\label{S_2_1}
The starting point for a formulation of the MEG problem is a mathematical model of the neural current densities that arise when
a large number of neurons in one or more small brain regions undergo a simultaneous discharge process. In mathematical terms,
this can be viewed as a set of point sources (i.e., current dipoles), each one representing one of the neuron populations.
To implement this model, the brain volume is discretized by $\nc$ cells, labelled by the index $c\@$.
Inside each cell, one reference point is chosen and identified by its position vector $\mathbf{r}\/(c)$ with respect to an appropriate reference system. A single current dipole is then allowed to be active at each point of the grid $\left\{\mathbf{r}\/(c)\right\}_{c\@=\/1}^\Nc\@$\vspace{1pt}: this requires considering, for all $c\@$, a current density  $\mathbf{p}\/(c)=\mathbf{q}\/(c)\,\delta\big(\mathbf{r}-\mathbf{r}\/(c)\big)\@$, where $\mathbf{q}\/(c)$ is an applied vector representing the moment of the dipole and the Dirac delta results from a limit process of a current density whose support concentrates into the point $\mathbf{r}(c)$. The moment $\mathbf{q}\/(c)\@$, in turn, can be expressed as the product of the unit vector $\mathbf{u}\/(c)$, which describes the direction of the dipole, and the strength $q\/(c)\@$, which can be either a real or a complex number, depending on the kind of topography considered\footnote{For instance, $q\/(c)\@$
is a complex number whenever the topography is obtained by selecting a specific frequency of the Fourier transform of the measured field.}.

Next, in order to formulate the equation linking the unknowns with the data of the MEG problem, the forward model of the head has to
be taken into account. This involves the computation, for each cell, of the so--called \emph{lead--field matrix} $\left[G(c)\right]^s_{\ k}\@$\vspace{1pt}, which is defined as follows \citep{haetal93}. Given a Cartesian orthonormal basis $\left\{\mathbf{e}_k\right\}_{k=1}^3\@$, consider a unit current dipole $\mathbf{e}_k\, \delta\big(\mathbf{r}-\mathbf{r}(c)\big)$ along each Cartesian direction:
the entry $\left[G(c)\right]^s_{\ k}$ is then defined as the measure, made by the $s\/$--th sensor, of the magnetic field produced by the dipole $\mathbf{e}_k\, \delta\big(\mathbf{r}-\mathbf{r}(c)\big)$. As a result, denoting by $\ns$ the total number of sensors, we get a family of $\nc$ matrices, each of them being of dimension $\@3 \times\ns$, which contains the information about the geometry and conductivity of the head model.

By referring the moment of a generic dipole to the basis $\left\{\mathbf{e}_k\right\}_{k=1}^3$, any dipole
can be represented as the linear combination
\begin{equation}
\mathbf{p}(c)=\mathbf{q}(c)\,\delta\big(\mathbf{r}-\mathbf{r}(c)\big)=
\sum_{k=1}^{3} q(c)\, u^k(c)\ \mathbf{e}_k\,\delta\big(\mathbf{r}-\mathbf{r}(c)\big).
\end{equation}

Let $\nd$ be the number of dipoles and, for $\nd\geq 1$, let $c\@(d)$ denote the cell where the $d\@$--th dipole is located. Then, by exploiting the linearity of the Biot-Savart equation \citep{haetal93}, the measure by
the $s\/$--th sensor of the magnetic field produced by the $\nd$ dipoles can be expressed as
\begin{equation}\label{federer}
 b^{\@s} = \sum_{d=1}^{\nd}\sum_{k=1}^3 \left[G\big(c\@(d)\big)\right]^s_{\ k}q\big(c\@(d)\big)\, u^{\/k}\big(c\@(d)\big) +
 \epsilon^{\@s},\ \ \ s=1,\ldots,\ns,
\end{equation}
being $\epsilon^{\@s}$ the value of the noise affecting the measurement. Whenever $\nd=0$, the sum on $d$ in
(\ref{federer}) is conventionally set to zero.

\medskip
According to Eq.\,\eqref{federer}, the inverse problem of MEG can be formalized as follows: given the array $\b:=(\@b^{\/1}, \ldots, b^{\@\Ns}\@)$ of the measurements of the magnetic field,
determine the state of the neural currents, namely the number $\nd$ of dipoles,  their positions (\/i.e.,
the cells $c\@(d)$ where they are located\@), the Cartesian components $u^k\left(c\@(d)\right)$ of their directions, and their strengths $q\left(c\@(d)\right)\@$.\\ \indent The analysis made above suggests defining the state space $\X$ of the neural currents as
\begin{equation}\label{eq_var_dim_mod}
  \X := \bigcup_{\nd=\@0}^{\nd^\text{\tiny{max.}}}\{\nd\}\times \X\/(\nd),
\end{equation}
being $\nd^\text{\tiny{max.}} (\@ \ll \nc)$ the maximum number of dipoles allowed by the model and  $\X\/(\nd)\@$ the state space for a fixed number $\nd$ of dipoles.
This is usually called a \emph{variable--dimension model} \citep{camory05, roca04}. A detailed definition of the space $\X\/(\nd)$ can be found in \ref{S_A_1}. Here, we merely point out the following two features:
\begin{enumerate}
  \item for the concept of  ``\@number of dipoles $\nd$\@'' to be well--posed, any ambiguous representation of a single dipole as a vector sum of dipoles applied at the same point should be avoided;
  \item as the \emph{order} in any $\nd\@$--ple  in $\X\/(\nd)$ has no actual physical meaning, any two $\nd\@$--ples differing only by a permutation of their components should be identified.
\end{enumerate}
A point in the space $\X$ shall be denoted by $x\@$ and consists in a pair $(\/N_D , j_{N_D}\/)$, each $j_{N_D}$ being an equivalence class of $N_D$--ples $(\/j^\narc{1}, \ldots, j^\narc{N_D}\/)$. In turn, for $d = 1, \ldots, N_D$, each $j^\narc{d}$ is of the form
 $\big(c^\narc{d},z^\narc{d},\varphi^\narc{d},q^\narc{d}\big)\@$, where $c^\narc{d}, q^\narc{d}$ are synonymous with $c\@(d),q\/(\/c\@(d)\/)$ above and $z^\narc{d},\varphi^\narc{d}$ are cylindrical coordinates which determine the components $u^k\/(c\@(d)\/)$ by means of
 Eq. \!\eqref{ziocil}.

\subsection{Statistical model}\label{S_2_3}
Generally speaking, the inverse problem of estimating neural currents from a single M\/E\/G topography is ill--posed because of 
non--detectable source configurations which make the kernel of the Biot--Savart operator non--trivial. We shall cope with this issue by resorting to a Bayesian approach, whereby the problem is recast in terms of statistical inference and the information content of any involved variable is coded into a probability distribution \cite{soka04}. In particular, the solution is represented by a posterior density $\pi\/(\/x\@|\@\b)\@$, combining the available information known before any measurement is made with the likelihood of the model, as expressed by Bayes theorem: $\@\pi\/(x\@|\@ \b) \propto p\/(x)\, \L\/(\b\@|\@x)\@$.

\subsubsection{Prior distribution.}
The parameters of the prior probability density $p\/(x)$ are set in order to reflect neurophysiological knowledge.
Moreover, as discussed above, no more than one dipole is allowed to be located in each cell. By Eq.\@\eqref{eq_var_dim_mod}, $\@p\/(x)\@$ can be decomposed into the product
\begin{equation*}
p\/(\nd)\,p_{\Nd}\/(j_{\Nd})\, =\, p\/(\nd)\ (\nd\@!)\ p_{\Nd}(j^{\narc{1}})\@ \cdots\, p_{\Nd}(j^{\narc{\nd}})\,,
\end{equation*} 
where\@:
\begin{itemize}
 \item $p_{\Nd}\/(j_{\Nd}):= p\@(j_{\Nd}\@|\,\nd)\@$;\vspace{3pt}
  \item $p(\nd)$ pertains to the number of sources. Since the use of a dipole model intrinsically entails an exiguous number of
  neural generators, we choose a Poisson distribution with a small parameter; simulations on synthetic data show that values around 
  $0.3$ provide good results;
  \\[-12pt]
  \item the factorial arises from the fact that the probability of the equivalence class $j_{\Nd}$ is here written as the sum of the probabilities of its representatives $\big(\@j^\narc{1},\ldots,j^\narc{\nd}\big)$;\vspace{1pt}
  \item each $p_{\Nd}(j^{\narc{d}})$ can itself be decomposed as
\begin{equation*}
 p_{\Nd}(j^{\narc{d}}) = p_{\Nd}\big(c^\narc{d}|c^\narc{1},\ldots,c^\narc{d-1}\big) \cdot p_{\Nd}(z^\narc{d}) \cdot p_{\Nd}(\varphi^\narc{d}) \cdot p_{\Nd}(q^{\narc{d}})\@.
\end{equation*}

As far as the single factors in the above equation are concerned, we make the following assumptions:
\begin{itemize}
\item since the cells have approximately the same extension, the prior distribution for the dipole location is uniform with respect to the set of vertices, 
namely $p_{\Nd}\left(c^{\narc{d}}|c^\narc{1},\ldots,c^\narc{d-1}\right) = \frac{1}{\plus70 \nc-(d-1)}$; 
\item the prior distributions $p_{\Nd}(z^\narc{d})\@$ and $p_{\Nd}(\varphi^\narc{d})\@$ for the dipole orientation
are uniform probability densities on the intervals $[0 , 1]$ and $[0, 2\/\pi)$ respectively. This amounts to uniformly sampling from the half--sphere \citep{shba96};
\item the prior distribution for the strength of the dipole moment  is log--uniform, i.e., $q^\narc{d} = \pm 10^{\@3 U}\@\cdot\@k\@$, 
where the sign has uniform distribution, $U$ is uniformly distributed in the interval $(\/0\/,\/1)\@$ and the  order of magnitude of the constant $k$ is chosen as
$10^{-10}$, so that the mean of the absolute value $\abs{q^\narc{d}}$ is a physiologically plausible value\@\footnote{\@The dipole strength 
is measured in $[\text{A}]\cdot[\text{m}]$ (in S.I. units), and its typical values are around $10^{-8}$.}.
\end{itemize}
\end{itemize}

\subsubsection{Likelihood function.} 
The analytical form of the likelihood is strictly related to the probability density function of the noise. 
According to our model, at each sensor $s=1,\ldots\ns$, the true value of the magnetic field generated by 
the neural currents is corrupted by a realization $\epsilon^{\@s}$ of the random variable $E^{\@s}$, as 
expressed by Eq.\@\eqref{federer}.
We assume each $E^{\@s}$ to be Gaussian, with zero mean value and variance $\sigma_{\text{\tiny noise}}^{\@2}\@$\vspace{1pt}, and not 
correlated with its counterparts at different sensors.  

This entails the likelihood to be the following multivariate Gaussian pdf:\vspace{3pt}
\begin{equation}\label{likelihoodt}
\L\@(\mathbf{b}\, |\, x)=\mathcal{N}_{\Ns}\left(\/\mathbf{b} - \sum_{d=1}^{\nd}\sum_{k=1}^3 \left[G\big(c\@(d)\big)\right]^{\@s}_{\ k}q\big(c\@(d)\big)\, u^{\/k}\big(c\@(d)\big)\@;\,\mathbf{0},\,\sigma_{\text{\tiny noise}}^{\@2}\@ \mathbf{I}_{\@\Ns\times\Ns}\/\right),
\end{equation}
where $\mathbf{I}_{\@\Ns\times\Ns}$ denotes the $\@\ns\times\ns\@$ identity matrix.

\section{An SMC for MEG source modeling}\label{S_3}
The posterior density described in the previous section is mostly a complicated function on a high--dimensional space, 
and is therefore an impractical solution to the problem.  In this sense, point estimates should better be computed and thus a numerical approximation of $\@\pi\/(\/x\@|\@\b)\@$ is needed.

In the next subsection we recall the key ingredients of SMC samplers, a recently developed class of methods for approximating a complex probability density of interest with a relatively small computational cost. The reader is referred to \cite{dedoja06} for a full description of the subject. In Sec. \!\!\ref{S_3_2} and Sec. \!\!\ref{S_3_3} we specify our choices concerning the implementation of the SMC sampler for the MEG inverse problem. An adaptation technique, providing better and faster results, is then described in Sec. \!\ref{S_3_4}. 
Finally, in Sec. \!\ref{S_3_5} we describe how point estimates of the dipole parameters can be computed from the approximated posterior
distribution.

\subsection{Sequential Monte Carlo Samplers}\label{S_3_1}
In most cases, $\@\pi\/(x\@|\@ \b)\@$ is complex enough to make direct importance sampling (\/IS\/) inapplicable. 
This is because any importance density will necessarily be too different from the target distribution, thus yielding a poor representation of the latter (see \cite{ru81}, p. 122).

Then, it becomes natural to resort to a sequential version of IS, whereby the solution is achieved gradually. To this end, a tempering sequence of distributions $\@\{\@\pi_i\/(x\/)\colon \X{}\to\R\,\}_{\@1\leq\@i\@\leq \ni}$
is built in such a way that $\@\pi_1\/(x):= p\/(x)\@$ be the prior distribution and $\pi_{\Ni}\/(x):= \pi\/(x\@|\@ \b)\@$ be the posterior distribution.  By construction, $\@\pi_1\/(x)\@$ can now be easily approximated by means of a sensible importance density $\eta_1\/(x)\@$; in our implementation, we shall make the choice $\eta_1\/(x):= p\/(x)\@$. For any $i$, the particle approximation $\@\{X^\narc{p}_{\@i}\@\}_{\plus62 p\@=1}^{\Np}\@$ of $\pi_i\/(x)\@$\vspace{1pt} is then moved to the subsequent step by means of a Markov kernel. However, it is typically impossible to compute the resulting importance weights analytically and therefore the standard IS estimate of $\pi_{i+1}\/(x)\@$
 is not available.
The SMC sampler technique overcomes this issue at the price of performing importance sampling in an increasingly larger space 
$\X^{\@i}\@ = \underbrace{\X \times \cdots \times \X}_{i \ \mathrm{ times}}$.  

Denoting by $x_{\@1:\@i}\@$ the point $(\/x_1, \ldots, x_i\/)\in \X^{\@i}\@$, an extended probability distribution $\tilde \pi_i\/(x_{\@1:\@i})\colon \X^{\@i} \to \R\@$ is introduced for any $i$, in such a way that its marginal with respect to $x_{1:\@i-1}$ coincides with $\pi_i\/(x_i)$. 
This can be done by introducing artificial backward Markov kernels $\{L_k\}_{\@1\leq\@k\@\leq \ni}$  
such that $\int_{\X} L_k\/(X_{k+1}, x_k)\@ d\/x_k = 1\@$ for any $k\@$, and setting 
\[
 \tilde \pi_i\/(x_{\@1:\@i}) := \pi_i\/(x_i)\@ \prod_{k = 1}^{i-1}\@ L_k\/(x_{k+1}, x_k)\@.
\]

On the other hand, the importance density $\eta_i\/(x_{1:\@i})$ on $\X^{\@i}\@$ is naturally obtained from
the initial distribution $\eta_1\/(x_1)$ by means of forward Markov kernels $\{K_k\}_{\@1\leq\@k\@\leq \ni}\@$, i.e., 
\[
 \eta_i\/(x_{1:\@i}) := \eta_1\/(x_1)\, \prod_{k=2}^i\@ K_k\/(x_{k-1}, x_k)\@.
\]

The corresponding importance weights are
\begin{equation}\label{eq_pesi1}
\begin{split}
 w_i\/(x_{1:\@i}) &:= \frac{\tilde \pi_i\/(x_{1:\@i})}{\eta_i\/(x_{1:\@i})} = 
\frac{\pi_i\/(x_i)\@ \prod_{k = 1}^{i-1}\@ L_k\/(x_{k+1}, x_k)}{\plus60 \eta_1\/(x_1)\, \prod_{k=2}^i\@ K_k\/(x_{k-1}, x_k)} = \\[3pt]
&= w_{i-1}\/(x_{1:\@i-1})\, \frac{\pi_i\/(x_i)}{\pi_{i-1}\/(x_{i-1})}\, \frac{L_{i-1}\/(x_i, x_{i-1})}{K_i\/(x_{i-1}, x_i)} =\\[3pt]
 &=: w_{i-1}\/(x_{1:\@i-1})\, \tilde w_i\/(x_{i-1}, x_i)\@,
\end{split}
\end{equation}
and therefore just the ``\@incremental weights\@'' $\tilde w_i\@$ are to be evaluated.

The construction above makes it possible to define, by induction, the following algorithm. Introduce the normalized importance weights $W_i^\narc{p} := w_i\/(X^\narc{p}_{1:\@i})\, /\, \sum_{j=1}^{\Np}\@ w_i\/(X^\narc{j}_{1:\@i})\@$; then, for $i= 1$, compute a particle approximation $\big\{X^\narc{p}_1, W^\narc{p}_1\big\}$ of $\pi_1\/(x_1)\@$. As already noticed, the weight function $w_1\/(x_1) = \pi_1\/(x_1)\@/\@\eta_1\/(x_1)\@$ can be computed exactly. 
Assuming that at iteration $i-1\@$ the particle approximation $\big\{X^\narc{p}_{1:\@i-1}, W^\narc{p}_{i-1}\big\}$ of $\tilde \pi_{i-1}\/(x_{1:\@i-1})$ is available, let the path of each particle evolve by means of the Markov kernel $K_i\/(x_{i-1}, x_i)$ and compute the incremental weight as in \eqref{eq_pesi1}.  The algorithm stops at $i = \ni$.

The variance of the (\@unnormalized\@) importance weights tends to increase with $i\@$ and this may lead to a very poor representation of the target density by the sample set. This phenomenon, called \emph{degeneracy}, is routinely measured by using the so--called Effective Sample Size (ESS), defined as $\text{ESS}\/\big(\{W_i^\narc{p}\}\big):= \big(\sum_{p\@=1}^\Np(\@W_i^\narc{p})^2\@\big)^{-1}$. If the degeneracy is too high (\@ for example $\text{ESS}\/\big(\{W_i^\narc{p}\}\big) < \np\@/\@2\@$\@), it is advisable to perform a resampling step.

\smallskip
In order to put the SMC sampler into practice, three ingredients are to be properly chosen: the sequence $\{\pi_i\}$, the transition kernels $\{K_i\}\@$ and the artificial backward transition kernels $\{L_i\}\@$. We shall discuss in the following subsections which choices are more suited to the MEG inverse problem. 

\subsection{Sequence of distributions}\label{S_3_2}
In a Bayesian framework, a natural choice for the sequence of artificial distributions is 
\begin{equation}\label{eq_sequenza}
  \pi_i := p\, \cdot\, \L^{\@f\/(i)}\,,\ \ \text{with}\ \,  f\/(i) := \sum_{k=1}^i\@ \delta_k\@;\ \, \delta_1\@:= 0\@;\ 
  \, \delta_{\Ni}\@:= 1 - \sum_{k=1}^{\Ni-1}\@ \delta_k\@;\ \, \delta_k\@\in (0,1) \@ \; \forall k \, . 
\end{equation}
We thus start from the prior distribution $p$ and go towards the posterior by increasing the exponent
of the likelihood function $\L$ with the iterations; intuitively, this choice corresponds to embodying the information
content of the data in the probability distributions step by step.

\smallskip
The specific choice of the increments $\delta_k$, appearing in the definition of the function $f\/(i)\@$, 
will be done in Sec. \!\ref{S_3_4}.
Here we just want to emphasize that Eq. \!\eqref{eq_sequenza} leads to an important consequence, which holds 
at least whenever the likelihood is a multivariate Gaussian density with covariance matrix $\sigma^2\@ \mathbf{I}\@$.
In this case, $\@\L^{f(i)}$ is still a multivariate Gaussian density with covariance matrix $\sigma_i^2\@ \mathbf{I}\@$, being $\sigma_i = \sigma / \sqrt{\plus20 f(i)}$.  This means that, as $i$ varies, the $\pi_i$'s can all be interpreted as posterior distributions for the unknown, differing from one another by the estimate of the noise standard deviation.
This sounds particularly interesting in those circumstances where the value 
of $\sigma$ is unknown, as the SMC provides for free a comparison between the different estimates corresponding to
different values of $\sigma$.

Moreover, such interpretation can be viewed
in connection with the theory of regularization for inverse problems: the sequence 
$\{ \log {\pi_i} \}$ closely reminds of the one--parameter family of functionals that underlie several 
regularization algorithms \citep{soka04}; $\sigma_i$ here plays the role of the regularization parameter, 
that tunes the balance between the penalty term (the prior) and the fit with the data (the likelihood).
In this connection, the SMC samplers not only provide an approximation to the full posterior,
rather than a single estimate as in regularization algorithms, but also explore a finite subset
of the so--called regularization path.

\subsection{Transition kernels}\label{S_3_3}
Within our variable--dimension model, states \ifaoas \linebreak \fi
 evolve by means of transition kernels $\{\@K_i\}$. We choose each $K_i$ to be a Reversible Jump MCMC kernel \citep{gr95} of invariant distribution $\pi_i\@$. More specifically, we shall assume 
\begin{equation}
 K_i\/(x_{i-1}, x_{i}) = K_i^\narc{b/d}\/(x_{i-1}, x')\ \cdot\ K_i^\narc{mov}\/(x',x_{i})\,,
\end{equation}
where:
\begin{itemize}
  \item $K_i^\narc{b/d}\@$ is a Reversible Jump Metropolis--Hastings (\/RJMH\/) kernel that accounts for a possible \emph{birth/death} move. One single birth is attempted with probability $P_\text{\tiny{birth}} = 1 / 3\@$; the location of the new dipole is uniformly distributed within the set of the (not already occupied) vertices, its orientation is uniformly distributed within $\mathds{P}^{\@2}\/(\R)$ and its strength has a log--uniform distribution. A death move is attempted with probability $P_\text{\tiny{death}} = 1 / 20\@$: in this case we propose to exclude one dipole, which is uniformly chosen 
  among the existing ones;
\item  $K_i^\narc{mov}\@$ is the part of the kernel which is in charge of the parameters evolution. 
We attempt to move each dipole separately and the possible changes in its location, orientation and strength are treated one at a time. This means that $K_i^\narc{mov}\@$ is actually the product of $3\@\cdot \nd\,$ Metropolis--Hastings (\/MH\/) kernels. As far as the proposals are concerned, we make the following assumptions. The new dipole location is drawn from a set of \textit{neighbours}, i.e., grid points within a radius of $1$\@cm, with probability proportional to a Gaussian centred at the starting position. 
The proposed orientation is obtained by means of  a zero--mean isotropic perturbation of the original one, implemented in Cartesian coordinates.  The new dipole strength is drawn from a Gaussian distribution of mean $q^\narc{d}$ and standard deviation $q^\narc{d} / 6$, so that there is negligible probability of going below $q^\narc{d} / 2\@$.  
\end{itemize}

\smallskip
As shown in \cite{dedoja06}, since $K_{i+1}$ is an MCMC kernel of invariant distribution $\pi_{i+1}$, we can define the backward kernel $L_i$ as
\begin{equation}
  L_{i}\/(x_{i+1}, x_{i}\@) := \frac{\pi_{i+1}\/(x_{i})\@K_{i+1}\/(x_{i}, x_{i+1})}{\pi_{i+1}\/(x_{i+1})}\@.
\end{equation}

By replacing this into Eq. \eqref{eq_pesi1}, we get
\begin{equation}\label{eq_agg_peso}
  \tilde w_{i}\/(x_{i-1}, x_{i}) = \pi_{i}\/(x_{i-1})\, /\, \pi_{i-1}\/(x_{i-1})\@.
\end{equation}

As a consequence, the weights $\{W_{i}^\narc{p}\}$  do not depend on the particles $\{X_{i}^\narc{p}\}$  and can thus be already computed at iteration $i-1$. This opens the possibility of resampling the approximation $\{W_{i}^\narc{p}, X_{i-1}^\narc{p}\}$ of $\pi_{i}\/(x_{i-1})\@$ before the $\{X_{i}^\narc{p}\}$ are sampled: heuristically, this enables to resample exactly those particles that are known to have a significant weight in the forthcoming iteration.

\subsection{Adaptive SMC}\label{S_3_4}
The choice of the increments $\delta_k$ in Eq. \!\eqref{eq_sequenza} is a key--point for a good performance of the whole algorithm: the sequence $\{\@\pi_i\/(x\/)\}_{\@1\leq\@i\@\leq \Ni}$ should come to a compromise between its smoothness (\@i.e., $\pi_i$ none too different from $\pi_{i+1}$\@)   and an acceptable computational cost.

We first tried some of the possible choices suggested in \cite{dedoja06}, such as different kinds of geometric paths. 
However, we observed that they lacked in homogeneity in the sense described above: the difference between two adjacent 
distributions was unavoidably too little in some phases of the algorithm while too high in others, thus often compromising 
the quality of results.

Therefore, in the present work, we choose the function $f\/(i)$ so that  the SMC sampler ``\/adapts\@'' to the data set. 
A similar technique is described in \cite{dedoja12}.

The construction grounds on the fact that, as explained above, our particular choice of the transition kernels makes it possible to evaluate the weights $w_{i+1}\/(x_{1:\@i+1})\@$ during iteration $i\@$. From Eqs. \eqref{eq_pesi1}, \eqref{eq_sequenza} and 	\eqref{eq_agg_peso}, we get
\begin{equation}
 w_{i+1}\/(x_{\/1:\@i+1}) = w_{i}\/(x_{1:\@i})\,\cdot\, \frac{\pi_{i+1}\/(x_{i})}{\pi_{i}\/(x_{i})}\ =\ 
w_{i}\/(x_{1:\@i})\, \cdot\, \L^{\@\delta_{i+1}}\,.
\end{equation}
We do not fix the value of $\delta_{i+1}$ beforehand but, on the contrary, let it vary within the closed set $[\@\delta_{\text{\tiny{min}}}\,,\,\delta_{\text{\tiny{max}}}\@]$. 
At each iteration, we first assume $\delta_{\/i+1} = \delta_{\text{\tiny{max}}}$
 and compute the ratio $\text{ESS}\/\big(\{W_{i+1}^\narc{p}\}\big)\, /\, \text{ESS}\/\big(\{W_i^\narc{p}\}\big)\@$, which is an indicator of how much $\pi_{i+1}$ will differ from $\pi_i\@$: if the value of such ratio falls into a fixed ``\/reliability interval\@'' $[\@I_{\text{\tiny{min}}}\,,\,I_{\text{\tiny{max}}}\@]\@$, we confirm the choice of $\delta_{\/i+1}\@$, otherwise we propose a new value by bisection and so on. 
 As a consequence, the total number of iterations $\ni$\vspace{1pt} is unknown at the beginning of the analysis, with the constraint that $\sum_{k=1}^{\Ni} \delta_k = 1\@$. 
 In our algorithm, we set the values $\delta_{\text{\tiny{min}}} = 10^{-5}\@$, $\delta_{\text{\tiny{max}}}= 10^{-1}\@$, $I_{\text{\tiny{min}}} = 0.9\@$ and
$I_{\text{\tiny{max}}} = 0.99\@$.

A schematic version of the algorithm is given below.

\begin{algorithm}
\caption{Adaptive SMC algorithm}\label{alg:asmc}
\begin{algorithmic}
\State{\textit{Initialization: sample the prior}}
\For{$p=1,\ldots,N_P$} 
 \State {draw $X_1^p$ from $p(x)$;}
\EndFor

\State Set $i=1$ and $f(1)=0$
\State{}
\State{\textit{Main cycle}}
\While{$f(i) \leq  1$} 
\State $i \rightarrow i+1$
\State{}
\State{\textit{Possible resampling step}}
\If{$ ESS(i) \leq \np/2 $}
\State {apply systematic resampling}
\EndIf
\State{}
\State{\textit{MCMC sampling}}
\For{$p=1,\ldots,N_P$} 
	\State{RJMH move: propose birth/death, then accept/reject}
 \For{$d=1,\ldots,N_D^p$}
 	\State{MH move for each dipole parameter: propose new value, then accept/reject}
 \EndFor
 \State compute tentative weights $\tilde{w}_{i+1}^p = \frac{\pi_{i+1}(X_{i}^p)}{\pi_{i}(X_{i}^p)}$
\EndFor

\State{}
\State{\textit{Normalize weights and compute Effective Sample Size}}
\State{\textbf{for} $p=1,\ldots,N_P$, \textbf{do} $w_i^p = \tilde{w}_i^p / W_i$, with $W_i = \sum_p \tilde{w}_i^p$ \textbf{end for}}
\State{Compute $ESS(i+1)$}

\State{}
\State{\textit{Adaptive determination of next exponent}}
\While{$ESS(i+1)/ESS(i) \geq 0.99 ||  ESS(i+1)/ESS(i) \leq 0.9$} 
	\State{increase/decrease $\delta_{i+1}$}
	\State{re-compute weights $w_i^p$ and $ESS(i+1)$}
\EndWhile
\State{Set $f(i+1) = f(i) + \delta_{i+1}$}

\EndWhile
\end{algorithmic}
\end{algorithm}

\subsection{Point Estimates}\label{S_3_5}
From the approximation to the artificial distributions provided by the ASMC sampler, point estimates at iteration $i$ are obtained as follows.
\begin{itemize}
  \item The estimated number of active sources $(\hat{N}_D)_{\@i}$ is the mode of the marginal distribution for the number
  of dipoles, which can be computed from the approximation to the $i$--th density as  
	  \begin{equation}
		P_{\@i}\/(\nd=k\@ |\@ \b)\ =\ \sum_{p\@=1}^\Np W^\narc{p}_{\@i}\ \delta\/\big(\@k, \@(\nd^\narc{p})_{\/i}\@\big)\,, 
		\label{post_N}
	  \end{equation}
being $\delta\/(\@\cdot\@,\cdot)\@$ the Kronecker delta function and $(\nd^\narc{p})_{\/i}$ the number of dipoles in the $p$--th particle at iteration $i$.\vspace{1pt}
  \item The estimated locations of the active sources are the $(\hat{N}_D)_{\@i}$ local modes of the intensity measure for the source locations, conditioned on the estimated number of sources and approximated as 
	\begin{equation}
	\label{intensity}
	P_{\@i}\/\big(c\@|\@\b , \hat{N}_D\big)\ =\ \sum_{p\@=1}^\Np W^\narc{p}_{\@i}\@ \delta\/\Big( (N_D^{(p)})_i, (\hat{N}_D)_{\@i} \Big) \Bigg(\@ \sum_{d\@=1}^{\text{\tiny $(N_D^{(p)})_{\@i}$}} 
        \delta\/\big(\@c,\@ c^\narc{d}\/(X^\narc{p}_{\@i} \@)\big)\@\Bigg)\,,        
	\end{equation}	
being $c^\narc{d}\/(X^\narc{p}_{\@i} \@)$ the location of the $d$--th dipole in the $p$--th particle at iteration $i$. The fact that only those particles with $(\hat{N}_D)_i$ dipoles contribute to estimating the source locations has been introduced in order to avoid mis--localization effects that may appear with model averaging.
  \item The direction and the intensity of the estimated dipoles are the mean values of the conditional distribution,
  conditioned on the source location and on the estimated number of dipoles, approximated as
  \begin{equation}
  \label{moment}
  \mathds{E}_{\@i}[\@\mathbf{q}\/(c)\@ |\@\b, \hat{N}_D\@] = \sum_{p\@=1}^\Np W^\narc{p}_{\@i}\@ \delta\/\Big( (N_D^{(p)})_i, (\hat{N}_D)_{\@i} \Big) \@ \Bigg(\@ \sum_{d\@=1}^{\text{\tiny $(N_D^{(p)})_{\@i}$}} 
        \mathbf{q}\/\big(c^\narc{d}\/(X^\narc{p}_{\@i} \@)\big)\, \delta\/\big(\@c,\@ c^\narc{d}\/(X^\narc{p}_{\@i} \@)\big)\@\Bigg)\,.
  \end{equation}
\end{itemize}

\section{Simulation experiments}\label{S_4}
Simulated data are used to validate and assess the performance of the proposed ASMC sampler.
Sec. \!\ref{S_4_1} describes how data are generated and which measures are used in order to appraise the discrepancy between the estimated and the true dipole configuration. The actual potential of the algorithm is then properly illustrated in Sec. \!\!\ref{S_4_2}, by means of an extensive treatment of the analysis of one single topography. Results coming from the entire amount of data sets are eventually summarized in Sec. \!\ref{S_4_3}.
\subsection{Data generation and discrepancy measures}\label{S_4_1}
We designed synthetic data with the purpose of investigating the behaviour of the method under a range of different experimental 
conditions, such as the number of sources, their configurations (\@i.e., diverse locations, orientations and strengths\@), and noise levels. 

The generation of the synthetic data was realized by using the same source grid and lead--field matrix that are used afterwards by the ASMC sampler. The geometry of the MEG device corresponds to that of a 306 channel Vectorview device, produced by Elekta Neuromag Oy, Helsinki, Finland. The computation of the lead--field matrix $\left[G(c)\right]^s_{\ k}\@$ has been carried out with a Boundary Element Method, starting from the geometry of the head of a real subject, extrapolated from MRI images with Freesurfer\@\footnote{\@ \texttt{http://surfer.nmr.mgh.harvard.edu}}. Source points are regularly spaced at $5$ mm distance. 

Overall we produced $1,200$ topographies subdivided into $100$ groups, in each of which  the number of sources $\nd$ and the noise level $\nu$ vary independently, with $\nd = 1, 2, 3, 4$ and $\nu = 0,\@ \nu_\text{\tiny \/low},\@ \nu_\text{\tiny \/high}$, for a total of $12$ different cases. The topography with $\nd$ dipoles is obtained from that with $\nd - 1$ sources just by adding the signal of the $\nd$--th source to the pre--existing one.

Dipole locations are uniformly drawn from the (\@unoccupied\@) brain grid points. Dipole intensities are set to 
$7$, $10$, $5$ and $8$\,nA$\cdot$m respectively. The orientations are chosen to be those producing the strongest signal at the specified location. This choice is motivated by the will to avoid patently ill-posed conditions. Indeed, it is well-known that any head geometry characterizes a field of directions such that, at each location, all dipoles orientated along it produce a negligible magnetic field compared to that generated by dipoles lying on its perpendicular plane; as an example, radial dipoles produce no magnetic field outside of a perfectly spherical conductor.

In the low--noise condition, the noise standard deviation is set to $5\%$ of the peak of the noise--free signal, corresponding to a Signal--to--Noise Ratio similar to that of evoked responses, while for the high noise level it is twice as large.

\smallskip
In the following, whenever the estimated quantities lack the subscript $i$, they are to be understood as evaluated at the final iteration 
$\ni$.

Now, let $\big(\/\h{N}_D, \h{j}_{\Nd}\/\big)$ and $\big(\/N_D, j_{\Nd}\/\big)$ be the estimated and the true dipole configuration respectively. In order to quantify the discrepancy between them, the following distances shall be used\@:
\begin{itemize}
\item $\Delta_{\Nd}$, which is the difference between the estimated and the true number of sources $\h{N}_D - \nd\@$;\vspace{1pt}	
\item $\Delta_{\@r}$, which quantifies the localization error. This is a non--trivial task when the estimated number of dipoles differs from the true one. Here we use a modified version of the OSPA metric \cite{scvovo08} with no penalty for cardinality errors, which are evaluated separately by $\Delta_{\Nd}$ above. 
 \begin{equation}
   \Delta_{\@r} :=
  \begin{cases}
   \min_{\pi \in \Pi_{\text{\tiny $\h{N}_D$}, \text{\tiny ${N}_D$}} } \frac{1}{\plus90 \h{N}_D}
   \sum_{d\@=1}^{\text{\tiny $\h{N}_D$}}\, \big|\@\r\/\big(\h{c}^{\narc{d}}\big) - \r\/\big(c^{\narc{\/\pi(d)\/}}\big)\big| & \mbox{if}\ \h{N}_D \leq \nd \,,\\[5pt]
\min_{\pi \in \Pi_{\text{\tiny ${N}_D$}, \text{\tiny $\h{N}_D$}} } \frac{1}{\plus9	0 \nd}
   \sum_{d\@=1}^{\Nd}\, \big|\@\r\/\big(\h{c}^{\narc{\/\pi(d)\/}}\big) - \r\/\big(c^{\narc{d}}\big)\big| & \mbox{if}\ \h{N}_D > \nd \, ,
  \end{cases}
  \end{equation}
  where $\Pi_{k,l}$ is the set of all permutations of $k$ elements drawn from
  $l$ elements. When $\h{N}_D = \nd$, $\Delta_{\@r}$ is the average distance between all the pairs of estimated and true source locations that is minimal with respect to all possible pairings between true and estimated sources. When $\h{N}_D \neq \nd$, the farther elements of
  the more numerous set are simply ignored, and $\Delta_{\@r}$ is computed as above.  
\end{itemize}

\subsection{Illustrative description of the output} \label{S_4_2}
\begin{figure}
\begin{center}
\subfloat[]
{\includegraphics[width=7cm,height=3.5cm]{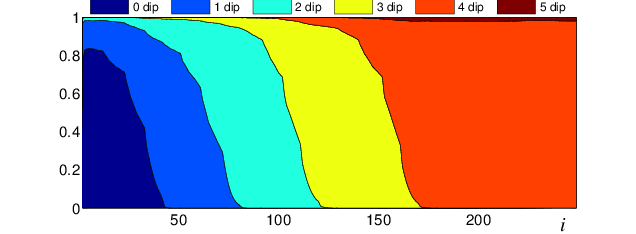}}
\qquad
\subfloat[]
{\includegraphics[width=7cm,height=3.5cm]{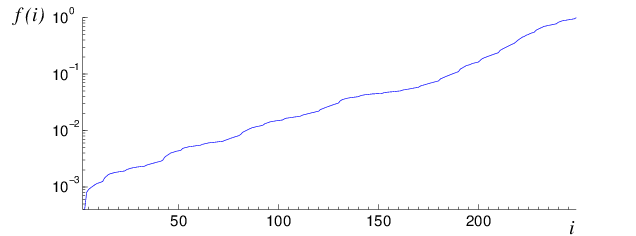}}
\end{center}
\caption{Behaviour of the SMC at different iterations: (a) the posterior probability $P_{\@i}\/(\nd\@ |\@ \b)\@$  and (b)%
the adaptively determined value of the exponent of the likelihood.}
\label{fig_synth_modsel}
\end{figure}

\begin{figure}
\begin{center}
{\includegraphics[width=12cm,height=4cm]{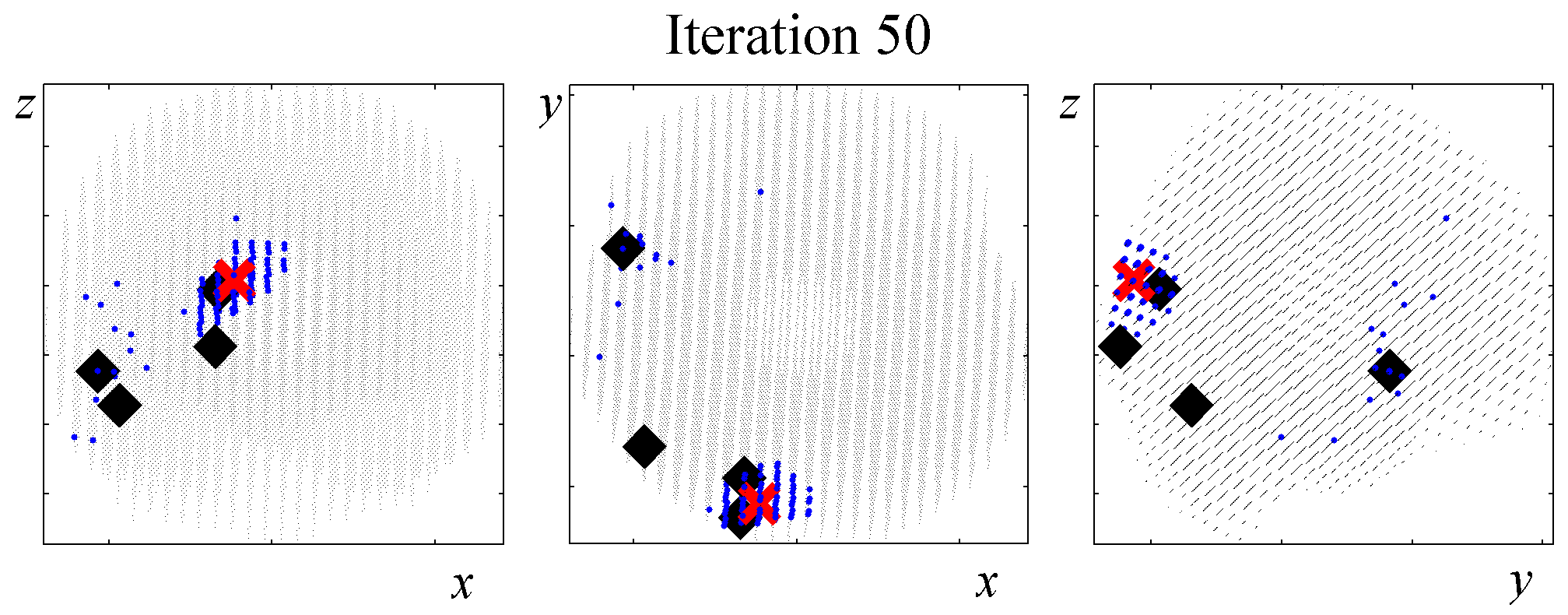}}
{\includegraphics[width=12cm,height=4cm]{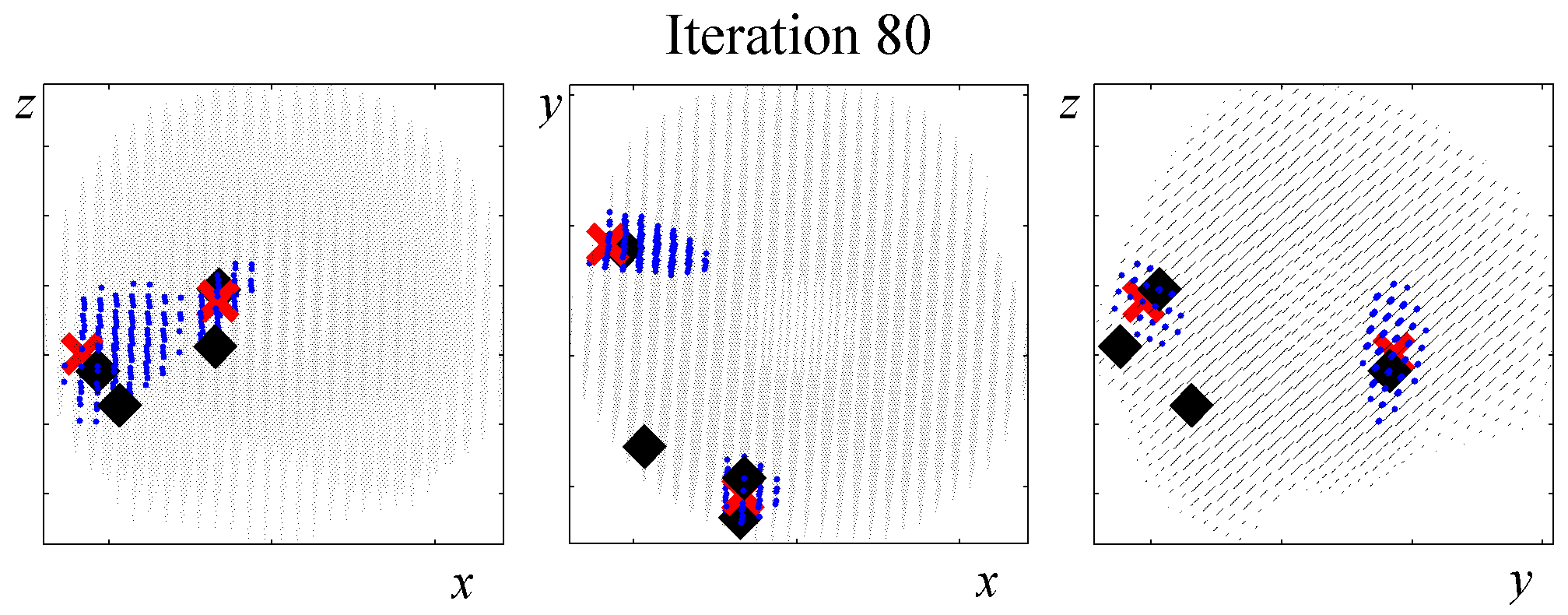}}
{\includegraphics[width=12cm,height=4cm]{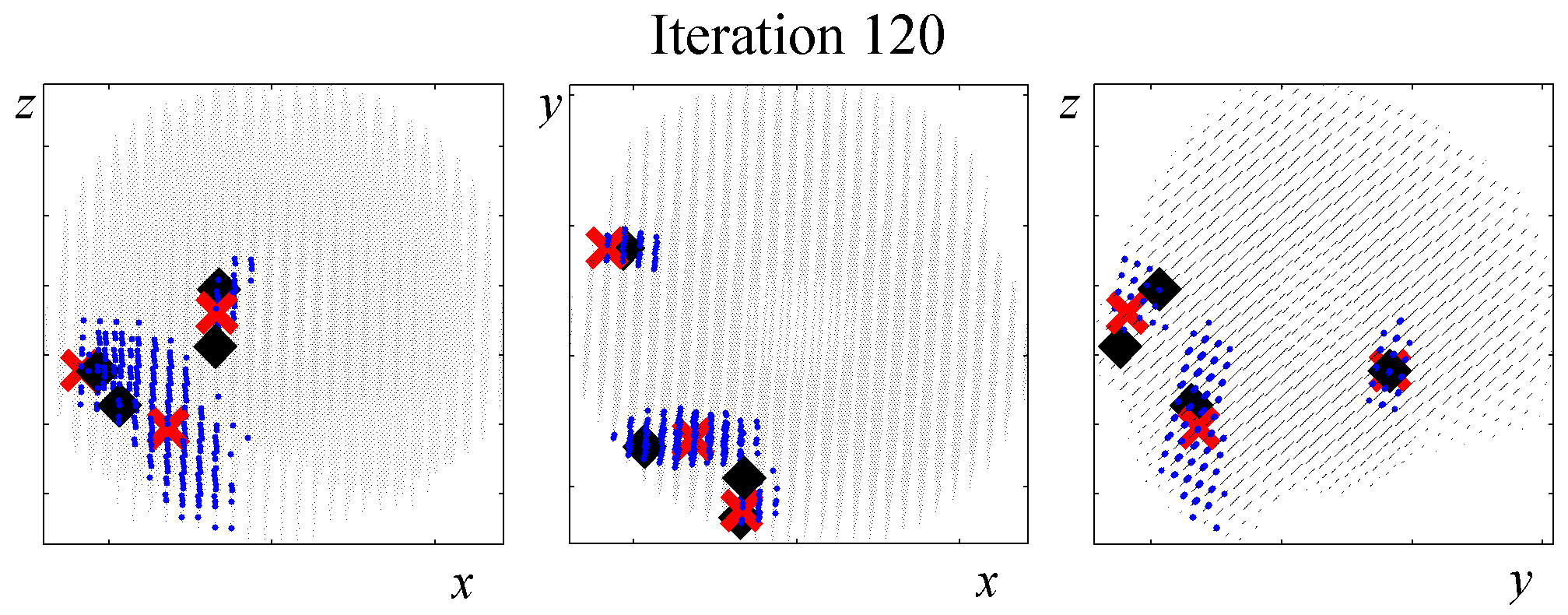}}
{\includegraphics[width=12cm,height=4cm]{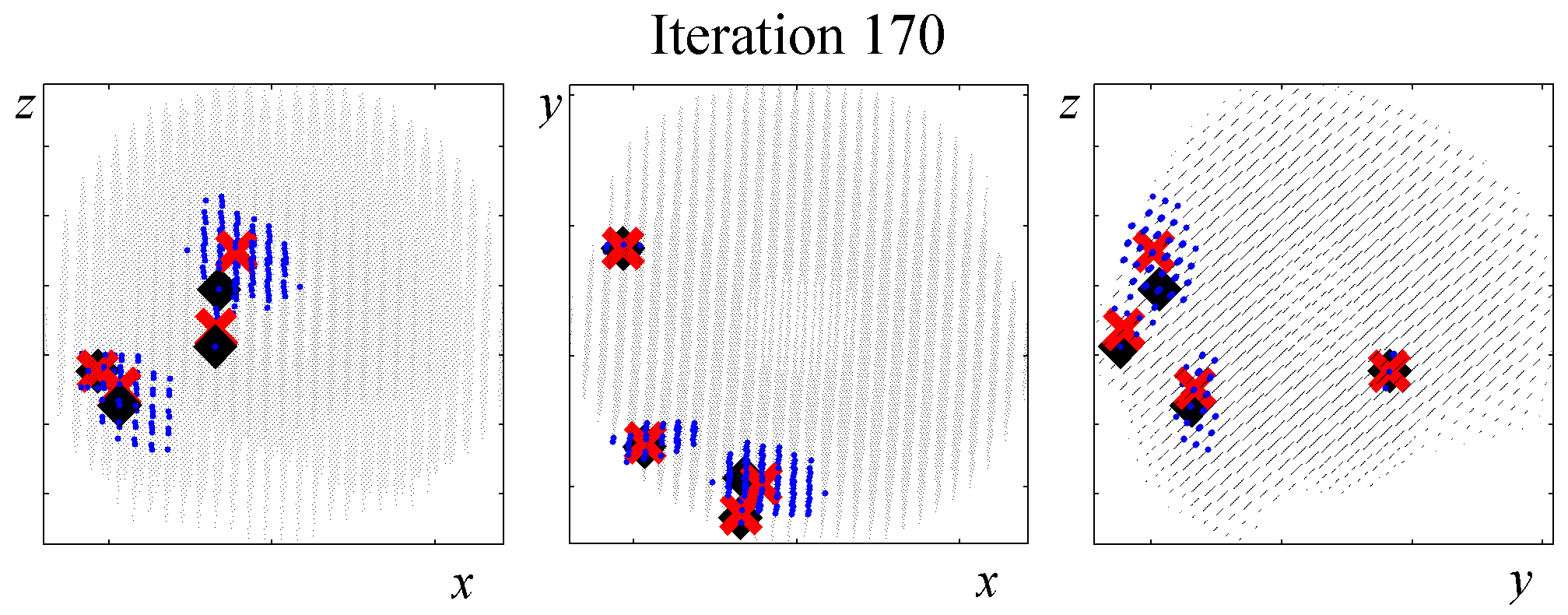}}
{\includegraphics[width=12cm,height=4cm]{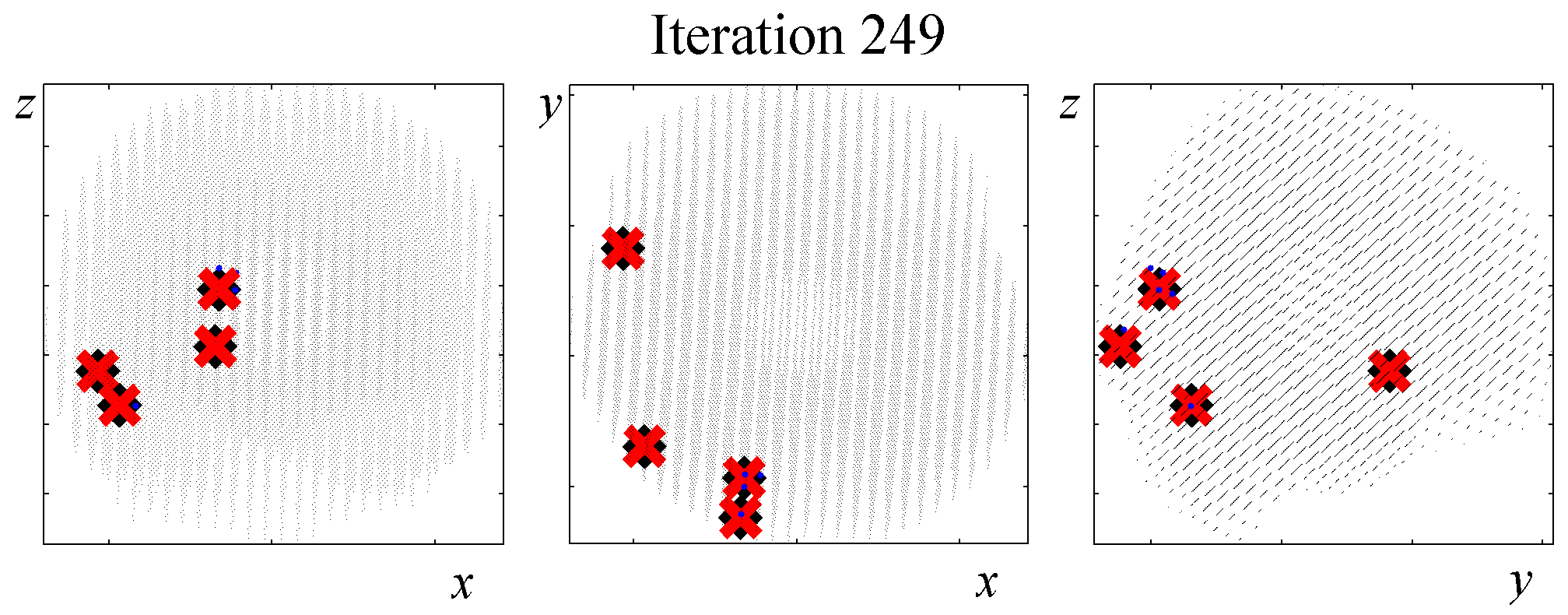}}
\end{center}
\caption{Coronal, axial and saggital views of the marginal probability for the dipole locations, taken at different iterations.}
\label{fig_synth_snapshots}
\end{figure}

In the present section, the analysis of a single synthetic topography is taken as an example so as to illustrate the behaviour of the algorithm. We consider a case where the known underlying source distribution is composed by four dipoles. 
Figure \ref{fig_synth_modsel} shows both the posterior probability $P_{\@i}\/(\nd\@ |\@ \b)\@$ for the number of sources and 
the (\/adaptively determined\/) value of the exponent in the likelihood as functions of the iterations. The posterior probability
mass smoothly moves from the zero--dipole scenario, which is preferred during the first iterations as the impact of the data
is still small, to models with ever--increasing dimension. Eventually, around iteration 170, the probability of the measured magnetic field being generated by four dipoles is close to one. 
The exponent of the likelihood grows almost exponentially with the iterations\@\footnote{\@Note the logarithmic scale on the vertical axis.}.
Figure \ref{fig_synth_snapshots} shows snapshots of the intensity measure for the source locations, at iterations $i = 50, 80, 120, 170, \ni$.
Black dots represent the  grid $\left\{\mathbf{r}\/(c)\right\}_{c\@=\/1}^\Nc\@$, while blue points represent those cells in which a dipole is located with intensity measure $P_{\@i}\/(c\@ |\@ \b, \hat{N}_D)\@$ exceeding $10^{-3}\@$. 
True dipole locations are denoted by black diamonds. Red crosses identify the estimated ones. 
The first three snapshots portray the situation at $i=50\@$ from the coronal, axial and saggital viewpoint respectively: the algorithm has localized the first dipole, the one producing the strongest signal.
Then, at $i=80\@$, a second source has also been recognized, while the uncertainty on the position of the first one has decreased.
It is worth to observe the peculiar shape of the marginal probability map for the more recently estimated dipole\/: 
the left--hand panel shows a wide uncertainty on both the $x$-- and the $z$--axis, while the middle panel indicates
that the uncertainty in the direction of the depth is much larger than in its orthogonal one;
this is indeed a well--known result in M\@E\@G literature \cite{haetal93}.
At subsequent iterations the probability maps tend to concentrate around the true source locations; eventually, in the last
iteration \ifarxiv \linebreak \fi 
(\@ which this time occurred for $i=249$\@), the estimated dipoles perfectly coincide with the true ones.

\subsection{Results}\label{S_4_3}
We analyzed the $1,200$ topographies described in Sec. \ref{S_4_1}; the value of $\sigma_{noise}$ in the likelihood function
was set equal to the standard deviation of noise, with a lower threshold of $ 10^{-14}$ fT/m, to avoid overly peaked posterior distributions
and bound the computational cost. We observed that moderate changes to the value of $\sigma_{noise}$ do not alter the results.
In all simulations the number of particles has been set to $\np=10,000$, which seemed to provide a reasonable compromise between the quality of the results and the computational cost.

For each topography we computed the point estimate
of the dipole configuration and calculated the discrepancy measures $\Delta_\Nd$ and $\Delta_r$ described above.
Average results along with standard deviations over the $100$ realizations are shown in Table \ref{table_error}\@. 

On average, the localization error as measured by $\Delta_r$ did not exceed $5\,$mm with the only exception of the high--noise / four--sources condition. It should be pointed out that the use of
the very same grid for both the data generation and the inverse method leads to a slightly underestimated localization error. In the
regular cubic grid used in this work, with a $5$ mm spacing, such underevaluation amounts to the average distance between a point and
the center of the cube the point belongs to, which is about $2.4$ mm.

Expectedly, the discrepancy between the estimated and the true configuration tends to increase with increasing noise.
This is due to two main effects. First, noise is a disturbance that may cause a small displacement of the peak of the posterior distribution;
as a consequence, even in the single-dipole condition the algorithm sometimes fail to localize the source exactly; but such average 
displacement is extremely small. Second and more important, a higher noise level leads to a decreased resolving power: when two
sources are very close to each other, it may become tolerable to replace them by a single source, whose dipole moment is the vector sum of the two. This is indeed reflected in the average $\Delta_{\Nd}$ becoming negative and decreasing with noise for the three--dipole and four--dipole configurations, where it is more likely that two sources placed at random in the brain are very close to each other.

The impact of the number of sources on the discrepancy is also clearly visible: the localization error increases
and $\Delta_{\Nd}$ becomes negative as the true sources grow in numbers. Even though these average values remain
well within the generally accepted localization errors (see for instance \cite{stetal02}, where a source is considered as correctly estimated if the localization error is below 20 mm), care must be taken in their interpretation. In fact, many three-dipole and four-dipole configurations are localized very well (see for instance the case shown in Figure \ref{fig_synth_snapshots}); in fewer cases, on the other hand, the estimated configuration can get rather far from the true configuration, thus contributing to increasing the average $\Delta_r$ as well as its standard deviation. 
However, this behaviour is not in general due to a poor particle approximation of the posterior distribution, but rather is mostly the consequence of two facts\/: the ill--posedness of the inverse problem and the limitations of the point estimation procedure. 

As far as the former aspect is concerned,  the presence of many dipoles may cause the data to be explained equally well by different configurations, which entails the multi--modality of the posterior distribution. This can be directly deduced from a visual 
inspection, e.g., of the marginal posterior distribution for the source locations, which shows a number of distinct peaks that is larger than the estimated number of dipoles. 

Secondly, point estimates are computed as local maxima of the marginal probability \eqref{intensity} and, for several reasons, 
they may not coincide with the true source locations; however, the true positions (\@or at least their 
immediate neighbours\@) are always assigned a non--negligible probability even when the estimated source configuration is clearly wrong. 
Moreover, a bad estimation performance in the 3-- or 4--dipoles scenario most often still yields a correct reconstruction of some of 
the sources (\@typically two\@).

\begin{table}
\begin{tabular}{||c||c|c|c|c||}
\hhline{|t:=:t:====:t|}
\backslashbox{$\nu$}{$\nd$}  & 1  & 2 & 3 & 4 \\
\hhline{|:=::====:|}
$\plus90 0$ &  $0.00 \pm 0.00$  & $0.01 \pm 0.10$ & $0.05 \pm 0.32$ & $-0.10 \pm 0.46$\\
 & $(0.0 \pm 0.0)\,$mm & $(0.8 \pm 3.1)\,$mm &$(2.4 \pm 4.4)\,$mm & $(4.6 \pm 5.3)\,$mm\\
\hline
$\plus90 \nu_{low}$  &  $0.00 \pm 0.00$  & $0.01 \pm 0.17$ & $-0.01 \pm 0.30$ & $-0.10 \pm 0.48$ \\
 & $(0.0 \pm 0.0)\,$mm & $(1.4 \pm 4.1)\,$mm & $(2.9 \pm 5.3)\,$mm & $(5.0 \pm 5.5)\,$mm\\
\hline
$\plus90 \nu_{high}$ & $0.00 \pm 0.00$ & $0.14 \pm 0.17$ & $-0.08 \pm 0.42$ & $-0.34 \pm 0.57$ \\
 & $(0.2 \pm 1.0)\,$mm & $(2.7 \pm 5.2)\,$mm & $(4.8 \pm 5.2)\,$mm & $(6.8 \pm 4.8)\,$mm\\
\hhline{|b:=:b:====:b|}
\end{tabular}

\vspace{5pt}
\caption{Discrepancy measures, averaged over 100 runs, for different noise conditions and different number of sources.
In each single box, $\Delta_{\Nd}$ (top) and $\Delta_{\@r}$ (bottom) are indicated.}
\label{table_error}
\end{table}

\section{Example on real data}\label{S_5}
To further test the ASMC sampler, we applied it to real data. Of course, direct validation is no longer possible,
since the ground truth is not known exactly.
To partially overcome this problem, we adopted the following strategy. 
Since the somatosensory response has been widely studied in the time domain \citep{maetal97} and is relatively well-understood,
we chose our single topographies to be single time points of the evoked response elicited by median nerve stimulation in a healthy subject. 
Moreover, we compared the results of the ASMC sampler with the source estimate obtained by applying the PF described in \cite{caetal11}
to the whole spatio--temporal recordings, and two widely used inverse methods, dynamic Statistical Parametric Mapping 
(dSPM, \cite{daetal00}) and sLORETA \citep{pa02}.

\subsection{Experimental Data}
Data from a Somatosensory Evoked Fields (SEFs) mapping experiment were acquired. The recordings were performed
at Istituto Neurologico Carlo Besta, Milano (Italy), by means of a $306$--channel MEG device (Elekta Neuromag Oy, Helsinki, Finland) 
comprising $204$ planar gradiometers and $102$ magnetometers in a helmet--shaped array. Informed consent and prior approval by the local 
ethics committee were obtained before the recording session. 
The position of the subject's head within the MEG helmet with respect to anatomical MRIs, obtained by using a $3$--Tesla MRI device 
(General Electric, Milwaukee, USA), was determined by means of a $3$\/D digitizer and four head position indicator coils.
The left median nerve at wrist was electrically stimulated at the motor threshold. 
The MEG signals were recorded at a sampling frequency of $2,000$ Hz and then band--pass filtered in the window $0.1-100$ Hz. 
Eye movements were monitored by Electrooculogram (EOG) in order to exclude artifacts from the MEG recordings: trials with EOG
or MEG exceeding $150$ mV or $3$ pT/cm, respectively, were neglected and 69 clean trials were averaged. 
The signal space separation method described in \citep{taetal04} was used to reduce external interference.

A thorough analysis of the neural response to the stimulation of the median nerve has been described in \cite{maetal97}, and
we take it as a reference here. In that study, the authors modelled the brain activity with multiple current dipoles; the number
of dipoles used to fit the response was established according to a subjective criterion, involving monitoring the difference between the
predicted and the measured data. Dipole parameters were fitted individually, one at a time, at user--selected time points. While 
some variability among subjects was observed, the SEF response was shown to obey to the following general scheme: 
first, the activation of the primary somatosensory cortex, in the hemisphere contralateral to the stimulation, starting about $20$ 
milliseconds after stimulus presentation; then, the activation of parietal sources, starting around $50$ milliseconds after the stimulus; 
finally, activity in frontal areas, after $100$ milliseconds.

\begin{figure}
\begin{center}
\includegraphics[width=16cm]{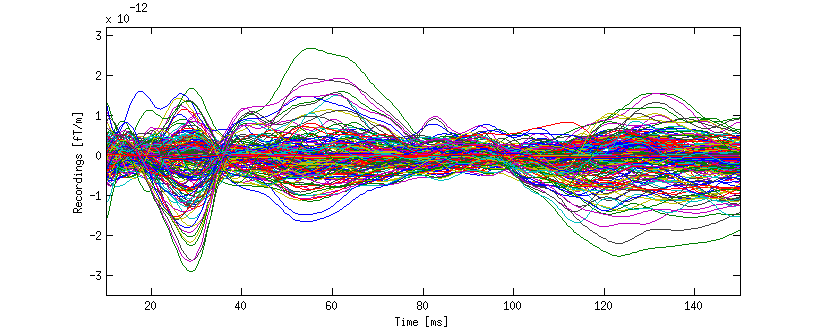}
\caption{Butterfly plot of the SEF data}
\label{fig:butterfly}
\end{center}
\end{figure}

The final, averaged data used in our study are plotted in Figure \ref{fig:butterfly}. 
The plot appears to confirm that there are three main temporal windows of interest: one between 20 and 35, one between 45 and 75, 
the last one starting around 110 milliseconds after stimulus.

\subsection{Results}
\begin{figure}
\begin{center}
\includegraphics[width=3.5cm]{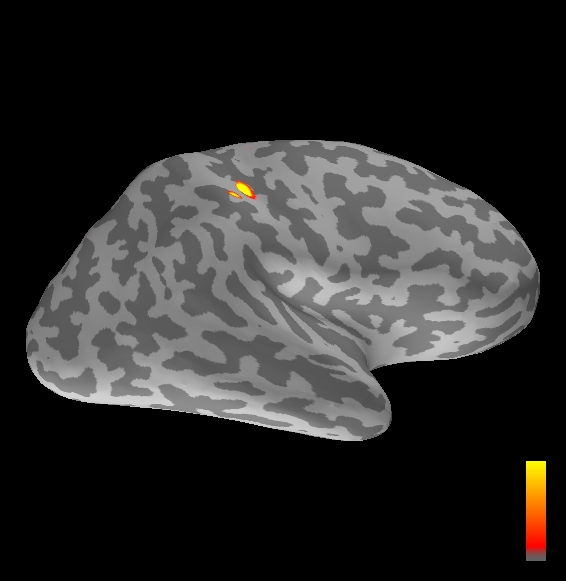}
\includegraphics[width=3.5cm]{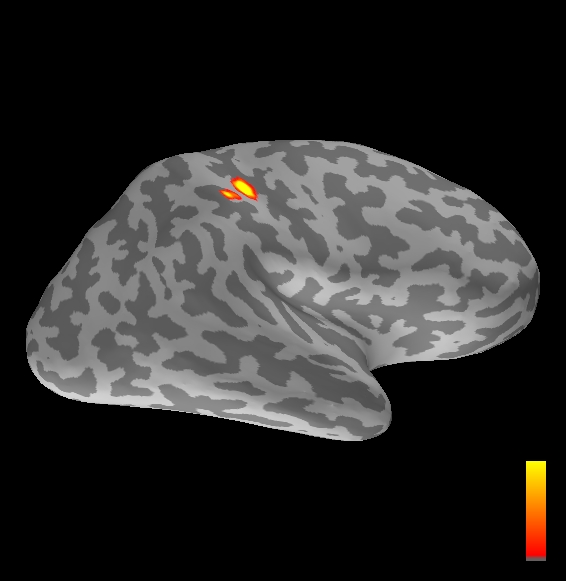}
\includegraphics[width=3.5cm]{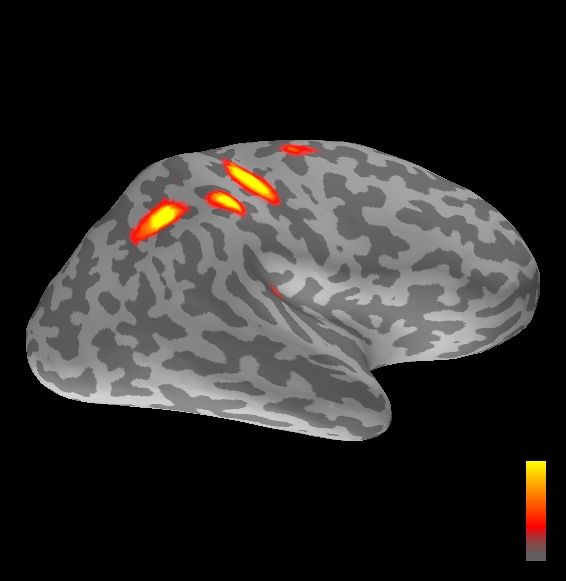}
\includegraphics[width=3.5cm]{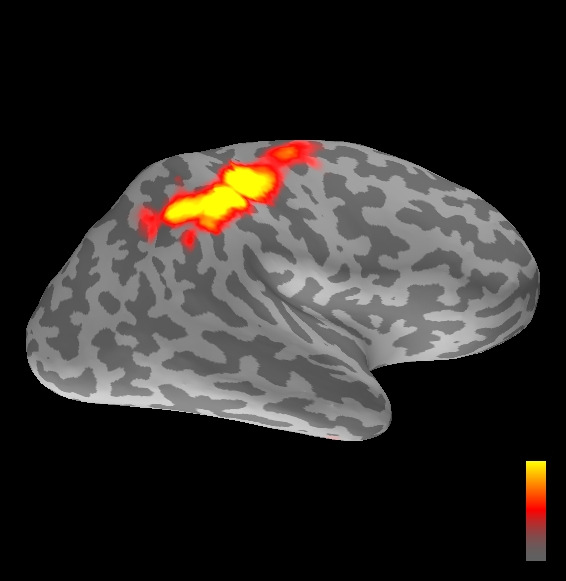}\\[4pt]
\includegraphics[width=3.5cm]{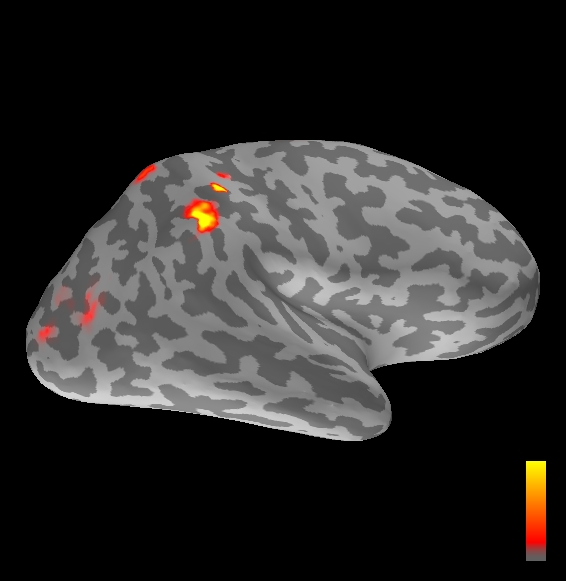}
\includegraphics[width=3.5cm]{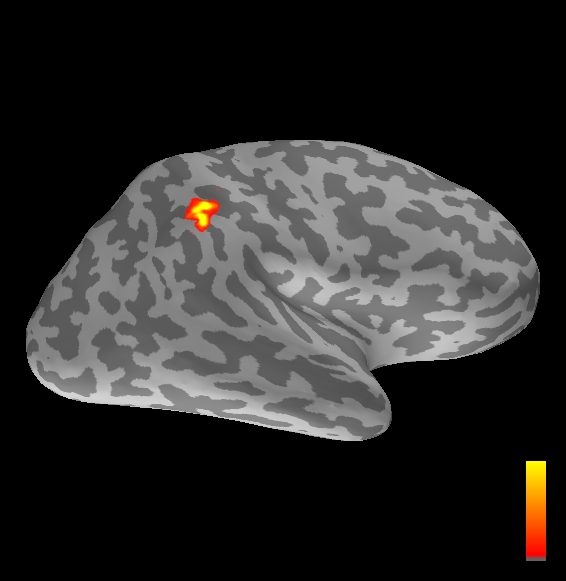}
\includegraphics[width=3.5cm]{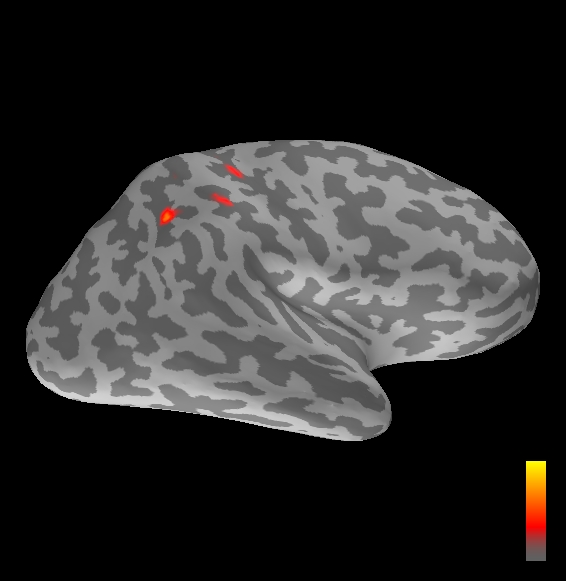}
\includegraphics[width=3.5cm]{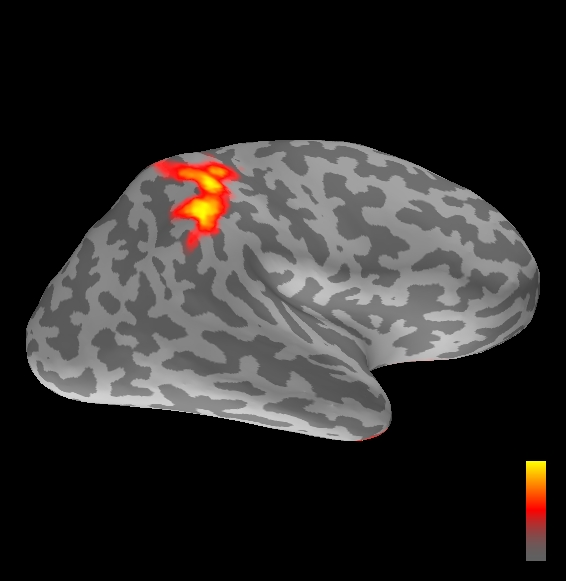}\\
\includegraphics[width=3.5cm]{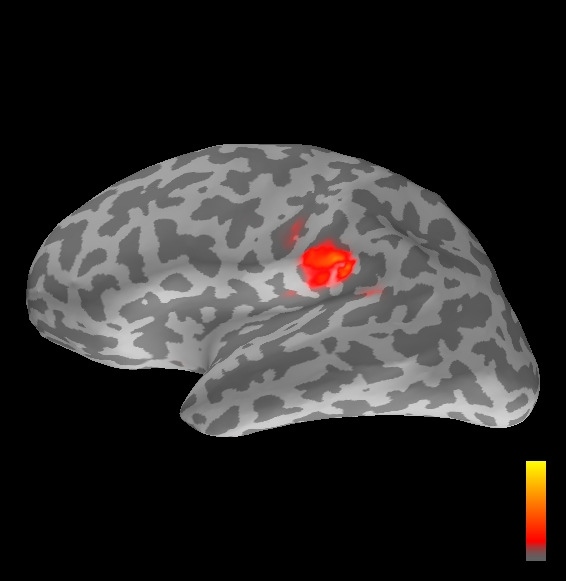}
\includegraphics[width=3.5cm]{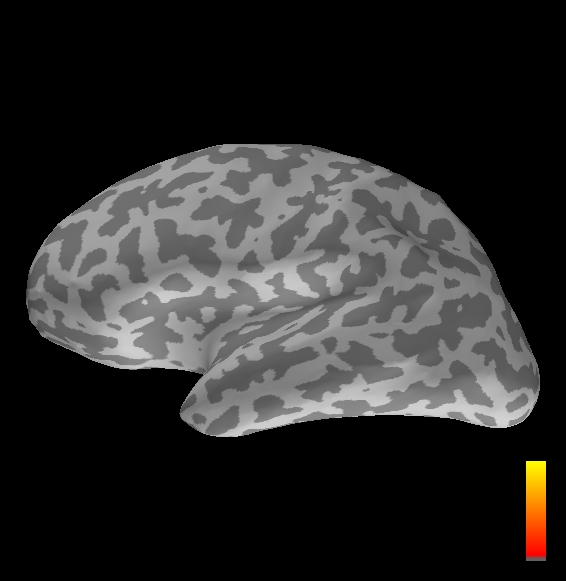}
\includegraphics[width=3.5cm]{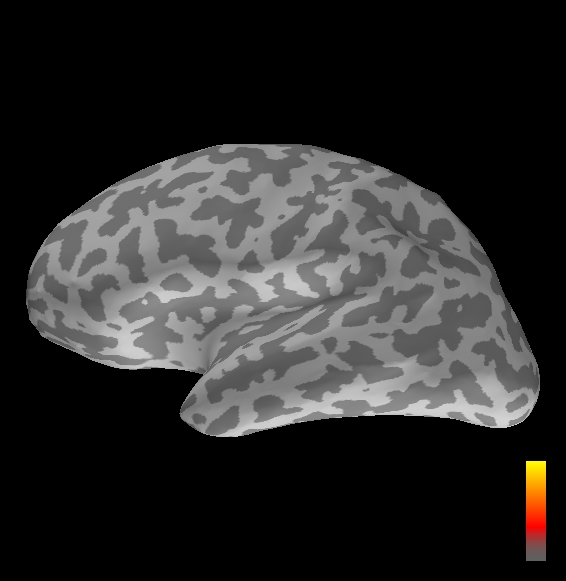}
\includegraphics[width=3.5cm]{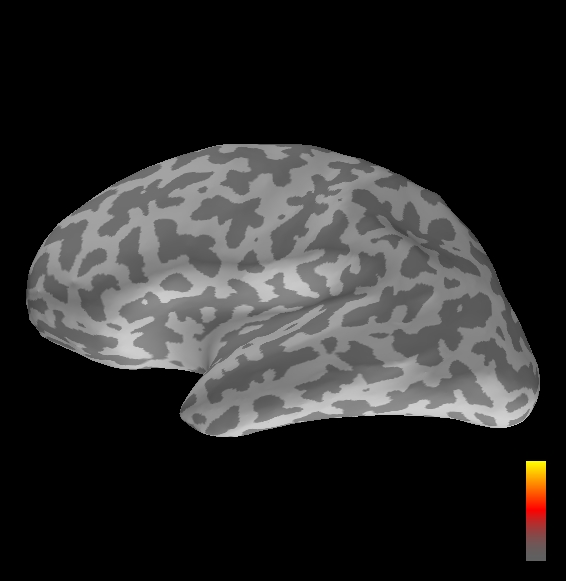}\\[4pt]
\includegraphics[width=3.5cm]{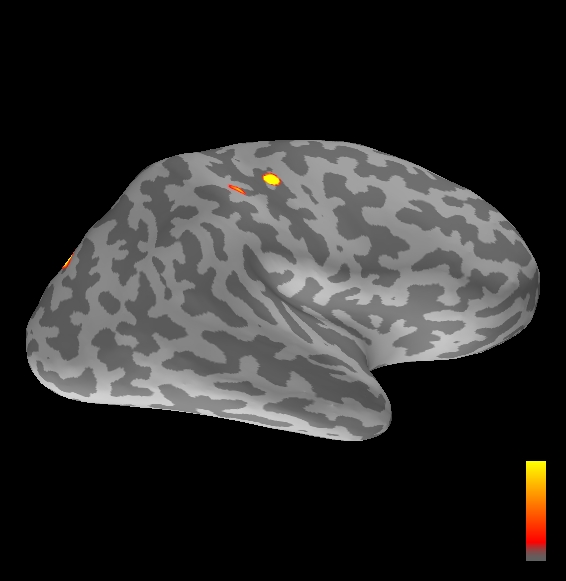}
\includegraphics[width=3.5cm]{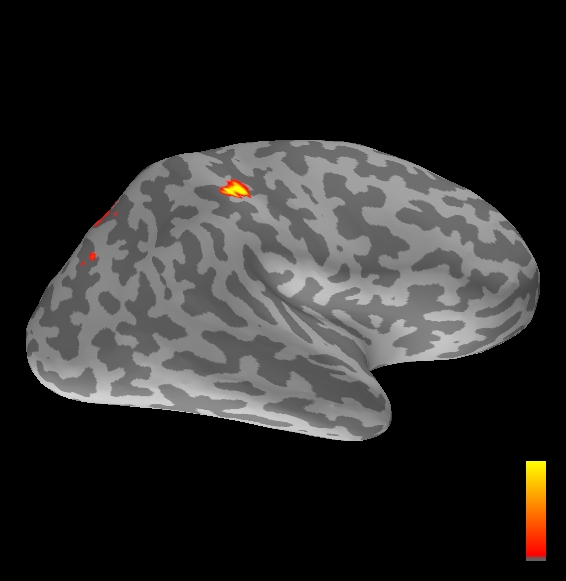}
\includegraphics[width=3.5cm]{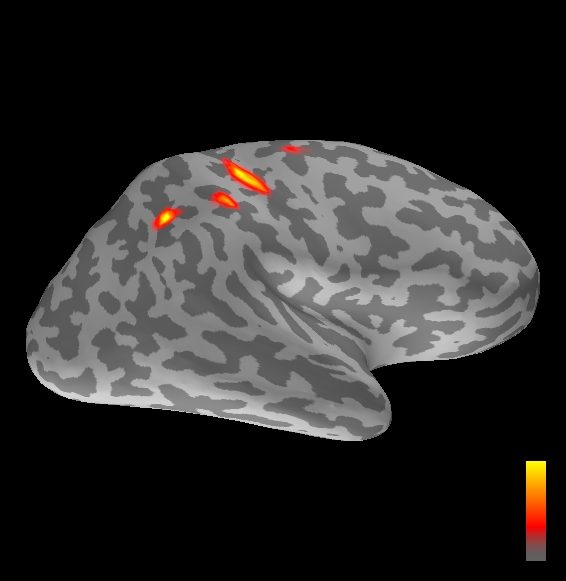}
\includegraphics[width=3.5cm]{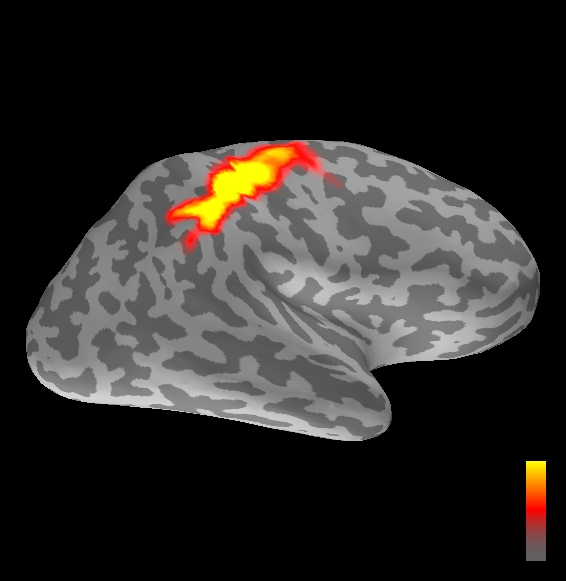}\\
\includegraphics[width=3.5cm]{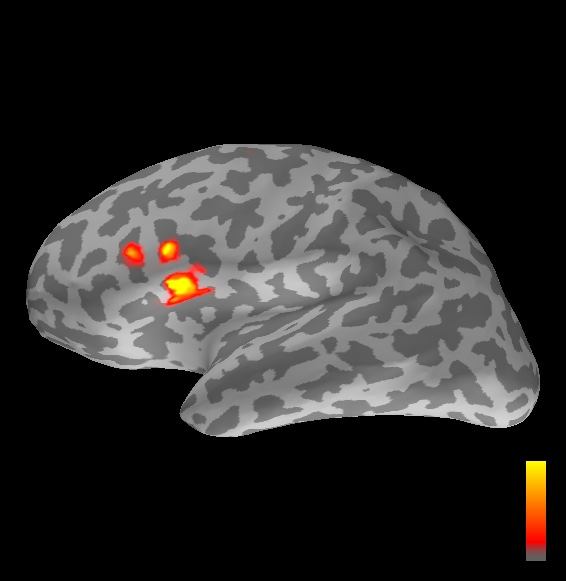}
\includegraphics[width=3.5cm]{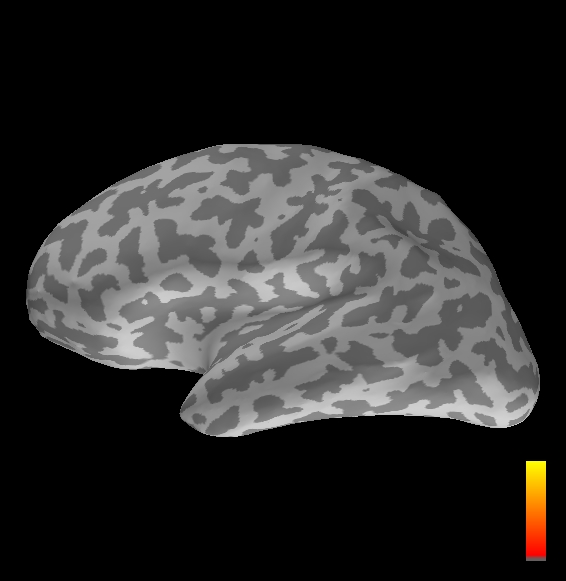}
\includegraphics[width=3.5cm]{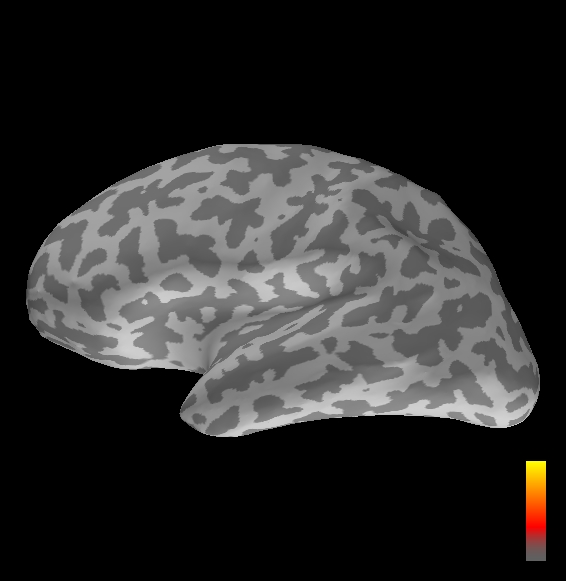}
\includegraphics[width=3.5cm]{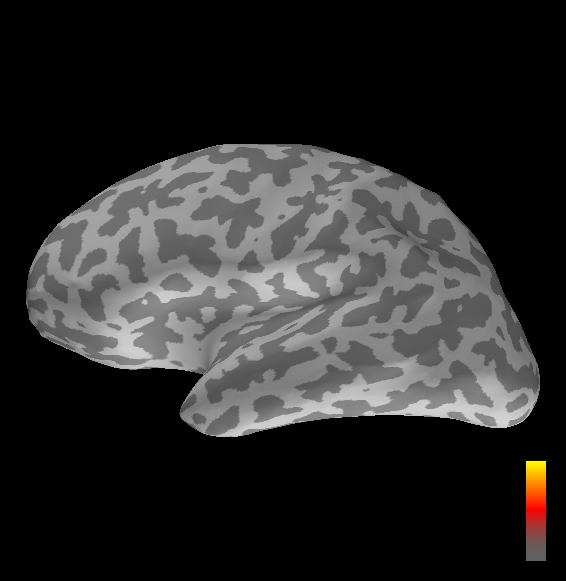}
\end{center}
\caption{Source estimates for the Somatosensory Evoked Fields by ASMC, PF, dSPM and sLORETA in the first, second, third and fourth column respectively. The first row represents $t = 25$ ms; the second and third rows portray the output at $t = 50$ ms in the right and left hemisphere respectively; the fourth and fifth rows do the same for $t = 120$ ms.}
\label{figure_SEF}
\end{figure}

We applied the ASMC sampler to MEG topographies taken
from the above recordings by selecting specific time points according to the previous analysis;
the parameter values in the algorithm were the same as those used for the analysis of synthetic data, with the only exception 
of the noise standard deviation $\sigma$, here estimated from the pre--stimulus interval.
In order to validate the results of the ASMC, we also computed source estimates using three other methods: 
a PF, that approximates the posterior distribution for the current dipoles conditioned
on the data \emph{up to} the selected time point; dSPM, which is based on a distributed source model with an $L^2$--prior, and 
consists in normalizing the Tikhonov regularized solution by the noise standard deviation; and sLORETA, which
is similar to dSPM but is claimed to have a smaller localization bias. 
Figure \ref{figure_SEF} shows the results at $t=30$, $50$ and $120$ ms after the stimulus onset. 
The results are visualized on a computer representation of the brain obtained by ``inflating'' the cortical surface: 
gray levels contain the anatomical information, light gray representing \emph{gyri} and dark gray representing \emph{sulci}; 
the activity estimate is coded in color scale, increasing from red to yellow. 
Importantly, this visualization allows activity in the sulci to be clearly visible; on the other hand, since neighbouring volumes may be moved
apart by the inflation process, distinct activity regions are often due to underlying volumetric masses that are very close to each other.

Before describing the results, let us comment on the qualitative difference between the images produced by 
the ASMC sampler and the PF, on the one hand, and those produced by dSPM and sLORETA, on the other.
First of all, we point out that all the quantities shown in the images of Fig. \ref{figure_SEF} are somehow related to the probability
of activation at specific locations. Indeed, for both the ASMC and the PF we plot the approximation of the intensity measure 
\eqref{intensity}; for any single grid point, this value can be interpreted as the probability of a dipole being at that location,
while it integrates, over a given volume $\Sigma$, to the mean number of dipoles within $\Sigma$. As for dSPM, the represented quantity
is a statistical value that is $t$--distributed under the null--hypothesis of zero activity; as a direct consequence, it also yields
a probability of activation, which is however not constrained to be dipolar. Similar considerations apply to sLORETA, although with a
different statistical distribution. Importantly, the representation of the results is clearly affected by the setting of the visualization 
threshold. Owing to the explained differences between the methods, it seems reasonable to use a different value for each method. At the same time, since the plotted quantity is a probability of activation, it seems right to use the same thresholding for different time points. 
In this connection, the thresholds in Fig. \ref{figure_SEF} have been chosen by hand following the guidelines just outlined.

Using the same thresholding and parameters at different time points makes the four methods respond differently
to the diverse intensities of the different sources.
Whenever a stronger source is active, both sLORETA and dSPM will tend to produce widespread estimates, while weaker sources
will be represented as small active areas. The behaviour of the ASMC and of the PF is the opposite: a stronger 
signal will lead to a precise localization of a dipolar source, and then to a focal marginal distribution 
for the location; a weaker signal will translate to higher uncertainty on the source position, and therefore a more widespread posterior map.

The phenomenon just described is indeed clearly visible in Figure \ref{figure_SEF}.
At ${t=25}$ ms, all the methods correctly identify the rather strong activation in the contra--lateral primary somatosensory 
cortex: the ASMC and the PF provide very focal maps, while dSPM and sLORETA provide compatible widespread estimates; 
dSPM also exhibits a more posterior peak which does not fit with the commonly agreed models of the response to median nerve 
stimulation; this may be due to the formerly described brain inflation.

At $t=50$ ms, the ASMC localizes two sources, one in the right hemisphere and a weaker one in the left hemisphere;
these sources appear to be compatible, for timing and location, with the Posterior Parietal Contra and Parietal Opercular
Ipsi described in \cite{maetal97}. The PF finds the very same source in the right hemisphere, but it estimates no dipoles in the
left one. Such discrepancy between these two methods, which are based on the same source model, is possibly due to two facts:
first, thanks to its iterative nature, the ASMC is likely to explore the state space more thoroughly and is therefore more suited
to identify weak sources; second, the log--uniform prior on the strength of the dipole moment in the ASMC has a wider
range than the corresponding Gaussian prior in the PF. Slightly different results are provided here by dSPM and sLORETA. The activity estimate computed by dSPM 
seems very similar to that of $t=25$ ms, having the same spatial distribution with a weaker intensity. 
The estimate of sLORETA in the right hemisphere includes that of the ASMC and of the PF. 
Neither dSPM nor sLORETA find significant activity in the left hemisphere.

At $t=120$ ms, the ASMC localizes again two sources: a stronger one in the right primary somatosensory cortex, and a weaker one 
in the left posterior frontal area. Both of them  seem to be compatible, for timing and location, with the results in \cite{maetal97},
the left dipole corresponding to the Frontal Ipsi source. Once again, the other three methods only recover the stronger source.

\section{Discussion}\label{S_6}
In the present paper, we described the application of an SMC sampler for the estimation of brain activity, modelled as multiple
current dipoles, from a single spatial distribution of magnetic field in MEG. 
We implemented an SMC sampler with a sequence of distributions built
by exponentiating the likelihood to an increasing value in the interval $\left[0,1\right]$. We made use of an adaptation 
technique that effectively tunes the speed of the algorithm, by monitoring the E\/S\/S at run time.

We applied the resulting ASMC sampler to a set of synthetic data, generated by source configurations containing up to four dipoles,
and affected by different levels of noise. We computed discrepancy measures between the true sources and the point estimates
provided by the approximated posterior distributions. The estimates are almost perfect for the one-- and two--dipole configurations,
and expectedly get worse with increasing noise in the three-- and four--dipole configurations. In all conditions, the average localization 
error remains below or slightly above $5$ mm, which is the grid spacing in our simulations. In addition, we observed that the fewer cases 
where results happen to be not completely satisfactory appear to
be more a consequence of the multi--modality of the distribution, and of the intrinsic limitations of any point estimation
technique, rather than pointing to an actual failure of the sampling process. 
Therefore, we suggest that further work is necessary to better investigate multi--modality.

We also tested the developed method against a set of somatosensory data recorded from a healthy subject.
We chose to use topographies taken from single time points, so as to compare the ASMC sampler with PF;
to further validate our analysis, we also computed source estimates using two well--known inverse methods, dSPM and sLORETA.
Results on experimental data suggest that the ASMC can correctly identify neural sources in real scenarios. All four methods
provided consistent estimates of the stronger sources; in addition, the ASMC finds tiny sources that are neglected by 
the other three methods. All the sources localized by the ASMC appear to be in accordance with the results of \cite{maetal97}, 
where the response to median-nerve stimulation is analyzed by means of multi--dipole models.

Further work is clearly necessary to better assess the performance of the proposed method in real scenarios.
For instance, it would be interesting to investigate the robustness of the solution with respect to (i) errors in the forward model and (ii) the number of particles used to approximate the posterior distribution.
Further, a systematic validation of the method with experimental data, including data in the frequency domain, is being carried on and will be the topic of a future publication.

From a mathematical viewpoint, one may ask whether and how it would be possible to exploit the linearity of the forward
equation with respect to the dipole moment, like it was done for the particle filter with Rao-Blackwellization \cite{caetal08}.
Preliminary results have been obtained in this direction, and they look encouraging.
In addition, the approach described in this paper could clearly be of interest in a number of different fields, where a sparse 
solution is desired and low--dimensional parametric models are available. Specifically, an application of the same methodology 
to an inverse problem in solar imaging \cite{huetal02} is currently ongoing.

More realistically, this work opens the possibility of applying Bayesian multi--dipole modeling to MEG data other than in the 
time domain.
We reckon that our ASMC sampler represents a valuable alternative to dipole fitting algorithms, with the considerable advantages
of providing an estimate of the number of sources and of not requiring careful initialization.
We hope that our efforts can contribute to provide novel, powerful and statistically sound analysis tools to the MEG community.


\appendix
\section{Multi--dipole state space}
\label{S_A_1}
The state space $\X$ of the neural currents has been defined in eq. \!\eqref{eq_var_dim_mod} as the variable dimension model
\[
 \X := \bigcup_{\nd=\@0}^{\nd^\text{\tiny{max.}}}\{\nd\}\times \X\/(\nd)\,.
\]

The spaces $\X\/(\nd)$ are, in turn, set out as follows.
First of all, $\X\/(\nd):=\emptyset$ for $\nd=0$. Then $\X\/(\nd):=\J$ for $\nd=1$, being $\@\J\/$ the state space  of a single dipole.

On account of Eq. \eqref{federer}, $\@\J\/$ can be defined as the Cartesian product $\N_{C}\times \mathbb{P}^2\/(\R)\times K$, where $\N_C:=[1,\nc]\cap \N\@$ is the space of cell numbers, the real projective plane $\mathbb{P}^2(\R)$ is the space of dipole directions and $K$ is the space of dipole strengths, which may be either $\R$ or $\C$; in the simulations of Sec. \ref{S_4} and Sec. \ref{S_5}, $K := \R\@$. We recall that $\mathbb{P}^2(\R)$
can be regarded as the quotient space of the $2$--sphere by identifying any two antipodal points. Referring $\R^3$ to cylindrical coordinates $(\rho,\varphi,z)\@$, the Cartesian components of a point $\mathbf{u}$ in $\mathbb{P}^2(\R)$ can thus be expressed as
\begin{equation}\label{ziocil}
 \left\{
\begin{array}{lcl}
  u^{1} & = & \sin(\arccos\/z)\@\cos\/\varphi \\
  u^{2} & = & \sin(\arccos\/z)\@\sin\/\varphi \\
  u^{3} & = & z,
\end{array}
\right.
\end{equation}
with $z\in [0,1]$ and $\varphi\in [0,2\pi)$ if $z\neq 0$, $\varphi\in [0,\pi)$ if $z=0$.
The two coordinates $z$ and $\varphi$, in the ranges just specified, are used to express the Cartesian components
(\ref{ziocil}) of the dipole direction $\mathbf{u}$. Summing up, the state of a single dipole is represented as the point
$j=(c,z,\varphi,q)\in\J$.

To address the case $\nd>1$, the first step is to consider the Cartesian product $\J^{\@\Nd}:=\J\times\ldots\times\J$ of $\nd$
copies of $\J$. A point of $\J^{\@\Nd}$ consists in a vector $\big(\@j^\narc{1},\ldots,j^\narc{\nd}\big)$ where,
for $d=1,\ldots,\nd\@$, each $\@j^\narc{d}\@$ is the quadruple $\big(c^\narc{d},z^\narc{d},\varphi^\narc{d},q^\narc{d}\big)\@$.
As pointed out in Sec. \ref{S_2_1}, care must be taken for the concept of  ``\@number of dipoles $\nd$\@'' to be well--posed: for this reason, the $\nd$ dipoles are constrained to be applied at different points. This entails subtracting from $\J^{\@\Nd}$ the set
\begin{equation}
\J^{\Nd}_{eq.}:=\left\{\big(j^\narc{1},\ldots,j^\narc{\nd}\big)\in\J^{\@\Nd}\,|\, \exists\, i,\,j\in\{1,\ldots,\nd\@\}\, :\, c\@(i)=c\@(j)\right\}
\end{equation}
to obtain the new space
$\J_{\Nd}:=\J^{\Nd}\setminus \J^{\Nd}_{eq.}$.
Moreover, since the ordering in any $\nd\@$--ple of $\J_{\Nd}$ if physically meaningless,
any two $\nd\@$--ples differing only by a permutation of their components $j^\narc{d}$ are identified. This amounts to setting onto $\J_{\Nd}$ the equivalence relation induced by the action of the symmetric group $S_{\Nd}$. The resulting quotient
space $\J_{\Nd} /\!\sim\@$ is eventually the appropriate choice for the space $\X\/(\nd)$. An equivalence class in $\J_{\Nd} /\!\sim\@$ is denoted by $\@j_{\Nd}\@$. Note that, for $\nd=0$, $\J_{\Nd} /\!\sim\ =\emptyset$ and, for $\nd=1$, $\J_{\Nd} /\!\sim\ =\J$, which is consistent with the previous settings.

As a result of the above investigation, the state space of neural currents is defined as the following disjoint union:
\begin{equation}\label{eq_spazio_stati}
\X:= \bigcup_{\nd=\@0}^{\nd^\text{\tiny{max.}}}\{\nd\}\times \J_{\Nd} / \sim\,.
\end{equation}


\section*{Acknowledgments}
The authors acknowledge Dr.\;Dunja Duran, Dr.\;Ferruccio Panzica, Dr.\;Davide Rossi, Fabio Rotondi and Dr.\;Elisa Visani (\/Istituto Neurologico Carlo Besta, Milano, Italy\/), who provided the 
experimental data used in Sec.\;\ref{S_5} and contributed to the neurophysiological interpretation of the results.
The authors also thank Prof.\;Michele Piana (\/Dipartimento di Matematica, Universit\`a di Genova, Italy\/), for
useful comments on the manuscript and support to this research.
A.S. kindly acknowledges Dr.\;Adam Johansen (\/Department of Statistics, University of Warwick, UK\/), for a gentle introduction 
and helpful discussions on the paper by Del Moral \emph{et al.} \cite{dedoja06} at the very beginning of this work.


\begin{thebibliography}{34}

\bibitem[\protect\citeauthoryear{Brookes et~al.}{2011}]{bretal11}
\begin{barticle}[author]
\bauthor{\bsnm{Brookes},~\bfnm{M.}\binits{M.}},
  \bauthor{\bsnm{Woolrich},~\bfnm{M.}\binits{M.}},
  \bauthor{\bsnm{Luckhoo},~\bfnm{H.}\binits{H.}},
  \bauthor{\bsnm{Price},~\bfnm{D.}\binits{D.}},
  \bauthor{\bsnm{Hale},~\bfnm{J.~R.}\binits{J.~R.}},
  \bauthor{\bsnm{Stephenson},~\bfnm{M.~C.}\binits{M.~C.}},
  \bauthor{\bsnm{Barnes},~\bfnm{G.~R.}\binits{G.~R.}},
  \bauthor{\bsnm{Smith},~\bfnm{S.~M.}\binits{S.~M.}} \AND
  \bauthor{\bsnm{Morris},~\bfnm{P.~G.}\binits{P.~G.}}
(\byear{2011}).
\btitle{Investigating the electrophysiological basis of resting state networks
  using magnetoencephalography}.
\bjournal{PNAS}
\bvolume{108}
\bpages{16783-16788}.
\end{barticle}
\endbibitem

\bibitem[\protect\citeauthoryear{Campi et~al.}{2008}]{caetal08}
\begin{barticle}[author]
\bauthor{\bsnm{Campi},~\bfnm{C.}\binits{C.}},
  \bauthor{\bsnm{Pascarella},~\bfnm{A.}\binits{A.}},
  \bauthor{\bsnm{Sorrentino},~\bfnm{A.}\binits{A.}} \AND
  \bauthor{\bsnm{Piana},~\bfnm{M.}\binits{M.}}
(\byear{2008}).
\btitle{A {R}ao-{B}lackwellized particle filter for magnetoencephalography}.
\bjournal{Inverse Problems}
\bvolume{24}
\bpages{025023}.
\end{barticle}
\endbibitem

\bibitem[\protect\citeauthoryear{Campi et~al.}{2011}]{caetal11}
\begin{barticle}[author]
\bauthor{\bsnm{Campi},~\bfnm{C.}\binits{C.}},
  \bauthor{\bsnm{Pascarella},~\bfnm{A.}\binits{A.}},
  \bauthor{\bsnm{Sorrentino},~\bfnm{A.}\binits{A.}} \AND
  \bauthor{\bsnm{Piana},~\bfnm{M.}\binits{M.}}
(\byear{2011}).
\btitle{Highly Automated Dipole EStimation}.
\bjournal{Computational Intelligence and Neuroscience}
\bvolume{2011}
\bpages{982185}.
\end{barticle}
\endbibitem

\bibitem[\protect\citeauthoryear{Capp{\'e}, Moulines and
  Ryd{\'e}n}{2005}]{camory05}
\begin{bbook}[author]
\bauthor{\bsnm{Capp{\'e}},~\bfnm{O.}\binits{O.}},
  \bauthor{\bsnm{Moulines},~\bfnm{E.}\binits{E.}} \AND
  \bauthor{\bsnm{Ryd{\'e}n},~\bfnm{T.}\binits{T.}}
(\byear{2005}).
\btitle{Inference in Hidden Markov Models}.
\bpublisher{Springer}.
\end{bbook}
\endbibitem

\bibitem[\protect\citeauthoryear{Chang, Ahlfors and Lin}{2013}]{chetal12}
\begin{barticle}[author]
\bauthor{\bsnm{Chang},~\bfnm{W.~T.}\binits{W.~T.}},
  \bauthor{\bsnm{Ahlfors},~\bfnm{S.~P.}\binits{S.~P.}} \AND
  \bauthor{\bsnm{Lin},~\bfnm{F.~H.}\binits{F.~H.}}
(\byear{2013}).
\btitle{Sparse current source estimation for MEG using loose orientation
  constraints}.
\bjournal{Human Brain Mapping}
\bvolume{34}
\bpages{2190-2201}.
\end{barticle}
\endbibitem

\bibitem[\protect\citeauthoryear{Dale et~al.}{2000}]{daetal00}
\begin{barticle}[author]
\bauthor{\bsnm{Dale},~\bfnm{A.}\binits{A.}},
  \bauthor{\bsnm{Liu},~\bfnm{A.~K.}\binits{A.~K.}},
  \bauthor{\bsnm{Fischl},~\bfnm{B.~R.}\binits{B.~R.}},
  \bauthor{\bsnm{Buckner},~\bfnm{R.~L.}\binits{R.~L.}},
  \bauthor{\bsnm{Belliveau},~\bfnm{J.~W.}\binits{J.~W.}},
  \bauthor{\bsnm{Lewine},~\bfnm{J.~D.}\binits{J.~D.}} \AND
  \bauthor{\bsnm{Halgren},~\bfnm{E.}\binits{E.}}
(\byear{2000}).
\btitle{Dynamic Statistical Parametric Mapping: Combining fMRI and MEG for
  High-Resolution Imaging of Cortical Activity}.
\bjournal{Neuron}
\bvolume{26}
\bpages{55-67}.
\end{barticle}
\endbibitem

\bibitem[\protect\citeauthoryear{Dassios, Fokas and Kariotou}{2005}]{dafoka05}
\begin{barticle}[author]
\bauthor{\bsnm{Dassios},~\bfnm{G.}\binits{G.}},
  \bauthor{\bsnm{Fokas},~\bfnm{A.~S.}\binits{A.~S.}} \AND
  \bauthor{\bsnm{Kariotou},~\bfnm{F.}\binits{F.}}
(\byear{2005}).
\btitle{On the non-uniqueness of the inverse {M}{E}{G} problem}.
\bjournal{Inverse Problems}
\bvolume{21}
\bpages{L1-L5}.
\end{barticle}
\endbibitem

\bibitem[\protect\citeauthoryear{{Del~Moral}, Doucet and
  Jasra}{2006}]{dedoja06}
\begin{barticle}[author]
\bauthor{\bsnm{{Del~Moral}},~\bfnm{P.}\binits{P.}},
  \bauthor{\bsnm{Doucet},~\bfnm{A.}\binits{A.}} \AND
  \bauthor{\bsnm{Jasra},~\bfnm{A.}\binits{A.}}
(\byear{2006}).
\btitle{Sequential {M}onte {C}arlo samplers}.
\bjournal{Journal of the Royal Statistical Society B}
\bvolume{68}
\bpages{411-436}.
\end{barticle}
\endbibitem

\bibitem[\protect\citeauthoryear{{Del~Moral}, Doucet and
  Jasra}{2012}]{dedoja12}
\begin{barticle}[author]
\bauthor{\bsnm{{Del~Moral}},~\bfnm{P.}\binits{P.}},
  \bauthor{\bsnm{Doucet},~\bfnm{A.}\binits{A.}} \AND
  \bauthor{\bsnm{Jasra},~\bfnm{A.}\binits{A.}}
(\byear{2012}).
\btitle{An adaptive sequential {M}onte {C}arlo method for approximate
  {B}ayesian computation}.
\bjournal{Statistics and Computing}
\bvolume{22}
\bpages{1009-1020}.
\end{barticle}
\endbibitem

\bibitem[\protect\citeauthoryear{Fokas, Kurylev and Marinakis}{2004}]{fokuma04}
\begin{barticle}[author]
\bauthor{\bsnm{Fokas},~\bfnm{A.~S.}\binits{A.~S.}},
  \bauthor{\bsnm{Kurylev},~\bfnm{Y.}\binits{Y.}} \AND
  \bauthor{\bsnm{Marinakis},~\bfnm{V.}\binits{V.}}
(\byear{2004}).
\btitle{The unique determination of neuronal currents in the brain via
  magnetoencephalography}.
\bjournal{Inverse Problems}
\bvolume{20}
\bpages{1067-1082}.
\end{barticle}
\endbibitem

\bibitem[\protect\citeauthoryear{Green}{1995}]{gr95}
\begin{barticle}[author]
\bauthor{\bsnm{Green},~\bfnm{P.~J.}\binits{P.~J.}}
(\byear{1995}).
\btitle{Reversible jump {M}arkov {C}hain {M}onte {C}arlo computation and
  {B}ayesian model determination}.
\bjournal{Biometrika}
\bvolume{82}
\bpages{711-732}.
\end{barticle}
\endbibitem

\bibitem[\protect\citeauthoryear{{H\"{a}m\"{a}l\"{a}inen}
  et~al.}{1993}]{haetal93}
\begin{barticle}[author]
\bauthor{\bsnm{{H\"{a}m\"{a}l\"{a}inen}},~\bfnm{M.}\binits{M.}},
  \bauthor{\bsnm{Hari},~\bfnm{R.}\binits{R.}},
  \bauthor{\bsnm{Knuutila},~\bfnm{J.}\binits{J.}} \AND
  \bauthor{\bsnm{Lounasmaa},~\bfnm{O.~V.}\binits{O.~V.}}
(\byear{1993}).
\btitle{Magnetoencephalography: theory, instrumentation and applications to
  non-invasive studies of the working human brain}.
\bjournal{Reviews of Modern Physics}
\bvolume{65}
\bpages{413-498}.
\end{barticle}
\endbibitem

\bibitem[\protect\citeauthoryear{Hurford et~al.}{2002}]{huetal02}
\begin{barticle}[author]
\bauthor{\bsnm{Hurford},~\bfnm{G.~J.}\binits{G.~J.}},
  \bauthor{\bsnm{Schmahl},~\bfnm{E.~J.}\binits{E.~J.}},
  \bauthor{\bsnm{Schwartz},~\bfnm{R.~A.}\binits{R.~A.}},
  \bauthor{\bsnm{Conway},~\bfnm{A.~J.}\binits{A.~J.}},
  \bauthor{\bsnm{Aschwanden},~\bfnm{M.~J.}\binits{M.~J.}},
  \bauthor{\bsnm{Csillaghy},~\bfnm{A.}\binits{A.}},
  \bauthor{\bsnm{Dennis},~\bfnm{B.~R.}\binits{B.~R.}},
  \bauthor{\bsnm{Johns-Krull},~\bfnm{C.}\binits{C.}},
  \bauthor{\bsnm{Krucker},~\bfnm{S.}\binits{S.}},
  \bauthor{\bsnm{Lin},~\bfnm{R.~P.}\binits{R.~P.}},
  \bauthor{\bsnm{McTiernan},~\bfnm{J.}\binits{J.}},
  \bauthor{\bsnm{Metcalf},~\bfnm{T.~R.}\binits{T.~R.}},
  \bauthor{\bsnm{Sato},~\bfnm{J.}\binits{J.}} \AND
  \bauthor{\bsnm{Smith},~\bfnm{D.~M.}\binits{D.~M.}}
(\byear{2002}).
\btitle{The {R}{H}{E}{S}{S}{I} Imaging Concept}.
\bjournal{Solar Physics}
\bvolume{210}
\bpages{61-86}.
\end{barticle}
\endbibitem

\bibitem[\protect\citeauthoryear{Hyv{\"a}rinen et~al.}{2010}]{hyetal10}
\begin{barticle}[author]
\bauthor{\bsnm{Hyv{\"a}rinen},~\bfnm{A.}\binits{A.}},
  \bauthor{\bsnm{Ramkumar},~\bfnm{P.}\binits{P.}},
  \bauthor{\bsnm{Parkkonen},~\bfnm{L.}\binits{L.}} \AND
  \bauthor{\bsnm{Hari},~\bfnm{R.}\binits{R.}}
(\byear{2010}).
\btitle{Independent component analysis of short-time Fourier transforms for
  spontaneousEEG/MEG analysis}.
\bjournal{NeuroImage}
\bvolume{49}
\bpages{257-271}.
\end{barticle}
\endbibitem

\bibitem[\protect\citeauthoryear{Jun et~al.}{2005}]{juetal05}
\begin{barticle}[author]
\bauthor{\bsnm{Jun},~\bfnm{S.~C.}\binits{S.~C.}},
  \bauthor{\bsnm{George},~\bfnm{J.~S.}\binits{J.~S.}},
  \bauthor{\bsnm{Par\'e-Blagoev},~\bfnm{J.}\binits{J.}},
  \bauthor{\bsnm{Plis},~\bfnm{S.~M.}\binits{S.~M.}},
  \bauthor{\bsnm{Ranken},~\bfnm{D.~M.}\binits{D.~M.}},
  \bauthor{\bsnm{Schmidt},~\bfnm{D.~M.}\binits{D.~M.}} \AND
  \bauthor{\bsnm{Wood},~\bfnm{C.~C.}\binits{C.~C.}}
(\byear{2005}).
\btitle{Spatiotemporal Bayesian inference dipole analysis for {M}{E}{G}
  neuroimaging data}.
\bjournal{NeuroImage}
\bvolume{28}
\bpages{84-98}.
\end{barticle}
\endbibitem

\bibitem[\protect\citeauthoryear{Mauguiere et~al.}{1997}]{maetal97}
\begin{barticle}[author]
\bauthor{\bsnm{Mauguiere},~\bfnm{F.}\binits{F.}},
  \bauthor{\bsnm{Merlet},~\bfnm{I.}\binits{I.}},
  \bauthor{\bsnm{Forss},~\bfnm{N.}\binits{N.}},
  \bauthor{\bsnm{Vanni},~\bfnm{S.}\binits{S.}},
  \bauthor{\bsnm{Jousmaki},~\bfnm{V.}\binits{V.}},
  \bauthor{\bsnm{Adeleine},~\bfnm{P.}\binits{P.}} \AND
  \bauthor{\bsnm{Hari},~\bfnm{R.}\binits{R.}}
(\byear{1997}).
\btitle{Activation of a distributed somatosensory cortical network in the human
  brain. {A} dipole modelling study of magnetic fields evoked by median nerve
  stimulation. Part {I}: location and activation timing of {S}{E}{F} sources}.
\bjournal{Electroencephalography and Clinical Neurophysiology}
\bvolume{104}
\bpages{281-289}.
\end{barticle}
\endbibitem

\bibitem[\protect\citeauthoryear{Pascual-Marqui}{2002}]{pa02}
\begin{barticle}[author]
\bauthor{\bsnm{Pascual-Marqui},~\bfnm{R.~M.}\binits{R.~M.}}
(\byear{2002}).
\btitle{Standardize Low resolution electromagnetic tomography
  (s{L}{O}{R}{E}{T}{A}: technical details}.
\bjournal{Methods and Findings in Experimental and Clinical Pharmacology}
\bvolume{24}
\bpages{5-12}.
\end{barticle}
\endbibitem

\bibitem[\protect\citeauthoryear{Robert and Casella}{2004}]{roca04}
\begin{bbook}[author]
\bauthor{\bsnm{Robert},~\bfnm{C.~P.}\binits{C.~P.}} \AND
  \bauthor{\bsnm{Casella},~\bfnm{G.}\binits{G.}}
(\byear{2004}).
\btitle{Monte Carlo Statistical Methods}, \bedition{Second} ed.
\bpublisher{Springer}.
\end{bbook}
\endbibitem

\bibitem[\protect\citeauthoryear{Rubinstein}{1981}]{ru81}
\begin{bbook}[author]
\bauthor{\bsnm{Rubinstein},~\bfnm{R.~Y.}\binits{R.~Y.}}
(\byear{1981}).
\btitle{Simulation and the Monte Carlo method}.
\bpublisher{John Wiley \& Sons Inc.}, \baddress{New York}.
\end{bbook}
\endbibitem

\bibitem[\protect\citeauthoryear{Salmelin}{2010}]{sa10}
\begin{bincollection}[author]
\bauthor{\bsnm{Salmelin},~\bfnm{R.}\binits{R.}}
(\byear{2010}).
\btitle{Multi-{D}ipole {M}odeling in {M}{E}{G}}.
In \bbooktitle{{M}{E}{G}: An introduction to methods}
\bpages{124-155}.
\bpublisher{Oxford University Press}.
\end{bincollection}
\endbibitem

\bibitem[\protect\citeauthoryear{Sarvas}{1987}]{sa87}
\begin{barticle}[author]
\bauthor{\bsnm{Sarvas},~\bfnm{J.}\binits{J.}}
(\byear{1987}).
\btitle{Basic mathematical and electromagnetic concepts of the biomagnetic
  inverse problem}.
\bjournal{Phys. Med. Biol.}
\bvolume{32}
\bpages{11-22}.
\end{barticle}
\endbibitem

\bibitem[\protect\citeauthoryear{Schuhmacher, Vo and Vo}{2008}]{scvovo08}
\begin{barticle}[author]
\bauthor{\bsnm{Schuhmacher},~\bfnm{D.}\binits{D.}},
  \bauthor{\bsnm{Vo},~\bfnm{B.~T.}\binits{B.~T.}} \AND
  \bauthor{\bsnm{Vo},~\bfnm{B.~N.}\binits{B.~N.}}
(\byear{2008}).
\btitle{A Consistent Metric for Performance Evaluation of Multi-Object
  Filters}.
\bjournal{IEEE Transactions on Signal Processing}
\bvolume{56}
\bpages{3447-3457}.
\end{barticle}
\endbibitem

\bibitem[\protect\citeauthoryear{Shao and Badler}{1996}]{shba96}
\begin{btechreport}[author]
\bauthor{\bsnm{Shao},~\bfnm{M.~Z.}\binits{M.~Z.}} \AND
  \bauthor{\bsnm{Badler},~\bfnm{N.~I.}\binits{N.~I.}}
(\byear{1996}).
\btitle{Spherical sampling by Archimedes' theorem}
\btype{Technical Report},
\binstitution{Department of Computer and Information Science, University of
  Pennsylvania}.
\end{btechreport}
\endbibitem

\bibitem[\protect\citeauthoryear{Somersalo and Kaipio}{2004}]{soka04}
\begin{bbook}[author]
\bauthor{\bsnm{Somersalo},~\bfnm{E.}\binits{E.}} \AND
  \bauthor{\bsnm{Kaipio},~\bfnm{J.~P.}\binits{J.~P.}}
(\byear{2004}).
\btitle{Statistical and computational inverse problems}.
\bpublisher{Springer Verlag}.
\end{bbook}
\endbibitem

\bibitem[\protect\citeauthoryear{Somersalo, Voutilainen and
  Kaipio}{2003}]{sovoka03}
\begin{barticle}[author]
\bauthor{\bsnm{Somersalo},~\bfnm{E.}\binits{E.}},
  \bauthor{\bsnm{Voutilainen},~\bfnm{A.}\binits{A.}} \AND
  \bauthor{\bsnm{Kaipio},~\bfnm{J.~P.}\binits{J.~P.}}
(\byear{2003}).
\btitle{Non-stationary magnetoencephalography by Bayesian filtering of dipole
  models}.
\bjournal{Inverse Problems}
\bvolume{19}
\bpages{1047-1063}.
\end{barticle}
\endbibitem

\bibitem[\protect\citeauthoryear{Sorrentino}{2010}]{so10}
\begin{barticle}[author]
\bauthor{\bsnm{Sorrentino},~\bfnm{A.}\binits{A.}}
(\byear{2010}).
\btitle{Particle {F}ilters for {M}agnetoencephalography}.
\bjournal{Archives of Computational Methods in Engineering}
\bvolume{17}
\bpages{213-251}.
\end{barticle}
\endbibitem

\bibitem[\protect\citeauthoryear{Sorrentino et~al.}{2009}]{soetal09}
\begin{barticle}[author]
\bauthor{\bsnm{Sorrentino},~\bfnm{A.}\binits{A.}},
  \bauthor{\bsnm{Parkkonen},~\bfnm{L.}\binits{L.}},
  \bauthor{\bsnm{Pascarella},~\bfnm{A.}\binits{A.}},
  \bauthor{\bsnm{Campi},~\bfnm{C.}\binits{C.}} \AND
  \bauthor{\bsnm{Piana},~\bfnm{M.}\binits{M.}}
(\byear{2009}).
\btitle{Dynamical {M}{E}{G} source modeling with multi-target Bayesian
  filtering}.
\bjournal{Human Brain Mapping}
\bvolume{30}
\bpages{1911-1921}.
\end{barticle}
\endbibitem

\bibitem[\protect\citeauthoryear{Sorrentino et~al.}{2013}]{soetal13}
\begin{barticle}[author]
\bauthor{\bsnm{Sorrentino},~\bfnm{A.}\binits{A.}},
  \bauthor{\bsnm{Johansen},~\bfnm{A.~M.}\binits{A.~M.}},
  \bauthor{\bsnm{Aston},~\bfnm{J.~A.~D.}\binits{J.~A.~D.}},
  \bauthor{\bsnm{Nichols},~\bfnm{T.~E.}\binits{T.~E.}} \AND
  \bauthor{\bsnm{Kendall},~\bfnm{W.~S.}\binits{W.~S.}}
(\byear{2013}).
\btitle{Dynamic filtering of static dipoles in {M}agnetoencephalography}.
\bjournal{Annals of Applied Statistics}
\bvolume{7}
\bpages{955-988}.
\end{barticle}
\endbibitem

\bibitem[\protect\citeauthoryear{Stam et~al.}{2009}]{stetal09}
\begin{barticle}[author]
\bauthor{\bsnm{Stam},~\bfnm{C.~J.}\binits{C.~J.}},
  \bauthor{\bsnm{{de~Haan}},~\bfnm{W.}\binits{W.}},
  \bauthor{\bsnm{Daffertshofer},~\bfnm{A.}\binits{A.}},
  \bauthor{\bsnm{Jones},~\bfnm{B.~F.}\binits{B.~F.}},
  \bauthor{\bsnm{Manshanden},~\bfnm{I.}\binits{I.}},
  \bauthor{\bsnm{{van~Cappellen~van~Walsum}},~\bfnm{A.~M.}\binits{A.~M.}},
  \bauthor{\bsnm{Montez},~\bfnm{T.}\binits{T.}},
  \bauthor{\bsnm{Verbunt},~\bfnm{J.~P.~A.}\binits{J.~P.~A.}},
  \bauthor{\bsnm{{de~Munck}},~\bfnm{J.~C.}\binits{J.~C.}},
  \bauthor{\bsnm{{van~Dijk}},~\bfnm{B.~W.}\binits{B.~W.}},
  \bauthor{\bsnm{Berendse},~\bfnm{H.~W.}\binits{H.~W.}} \AND
  \bauthor{\bsnm{Scheltens},~\bfnm{P.}\binits{P.}}
(\byear{2009}).
\btitle{Graph theoretical analysis of magnetoencephalographic functional
  connectivity in {A}lzheimer’s disease}.
\bjournal{Brain}
\bvolume{132}
\bpages{213-224}.
\end{barticle}
\endbibitem

\bibitem[\protect\citeauthoryear{Stenbacka et~al.}{2002}]{stetal02}
\begin{barticle}[author]
\bauthor{\bsnm{Stenbacka},~\bfnm{L.}\binits{L.}},
  \bauthor{\bsnm{Vanni},~\bfnm{S.}\binits{S.}},
  \bauthor{\bsnm{Uutela},~\bfnm{K.}\binits{K.}} \AND
  \bauthor{\bsnm{Hari},~\bfnm{R.}\binits{R.}}
(\byear{2002}).
\btitle{Comparison of Minimum Current Estimate and Dipole Modeling in the
  Analysis of Simulated Activity in the Human Visual Cortices}.
\bjournal{NeuroImage}
\bvolume{16}
\bpages{936-943}.
\end{barticle}
\endbibitem

\bibitem[\protect\citeauthoryear{Stoffers et~al.}{2007}]{stetal07}
\begin{barticle}[author]
\bauthor{\bsnm{Stoffers},~\bfnm{D.}\binits{D.}},
  \bauthor{\bsnm{Bosboom},~\bfnm{J.~L.~W.}\binits{J.~L.~W.}},
  \bauthor{\bsnm{Deijen},~\bfnm{J.~B.}\binits{J.~B.}},
  \bauthor{\bsnm{Wolters},~\bfnm{E.~C.}\binits{E.~C.}},
  \bauthor{\bsnm{Berendse1},~\bfnm{H.~W.}\binits{H.~W.}} \AND
  \bauthor{\bsnm{Stam},~\bfnm{C.~J.}\binits{C.~J.}}
(\byear{2007}).
\btitle{Slowing of oscillatory brain activity is a stable characteristic of
  Parkinson’s disease without dementia}.
\bjournal{Brain}
\bvolume{130}
\bpages{1847-1860}.
\end{barticle}
\endbibitem

\bibitem[\protect\citeauthoryear{Sutherland and Tang}{2006}]{suta06}
\begin{barticle}[author]
\bauthor{\bsnm{Sutherland},~\bfnm{M.~T.}\binits{M.~T.}} \AND
  \bauthor{\bsnm{Tang},~\bfnm{A.~C.}\binits{A.~C.}}
(\byear{2006}).
\btitle{Reliable detection of bilateral activation in human primary
  somatosensory cortex by unilateral median nerve stimulation}.
\bjournal{NeuroImage}
\bvolume{33}
\bpages{1042-1054}.
\end{barticle}
\endbibitem

\bibitem[\protect\citeauthoryear{Taulu, Kajola and Simola}{2004}]{taetal04}
\begin{barticle}[author]
\bauthor{\bsnm{Taulu},~\bfnm{S.}\binits{S.}},
  \bauthor{\bsnm{Kajola},~\bfnm{M.}\binits{M.}} \AND
  \bauthor{\bsnm{Simola},~\bfnm{J.}\binits{J.}}
(\byear{2004}).
\btitle{Suppression of interference and artifacts by the signal space
  separation method}.
\bjournal{Brain Topography}
\bvolume{4}
\bpages{269-275}.
\end{barticle}
\endbibitem

\bibitem[\protect\citeauthoryear{Uda et~al.}{2012}]{udetal12}
\begin{barticle}[author]
\bauthor{\bsnm{Uda},~\bfnm{T.}\binits{T.}},
  \bauthor{\bsnm{Tsuyuguchi},~\bfnm{N.}\binits{N.}},
  \bauthor{\bsnm{Okumura},~\bfnm{E.}\binits{E.}},
  \bauthor{\bsnm{Sakamoto},~\bfnm{S.}\binits{S.}},
  \bauthor{\bsnm{Morino},~\bfnm{M.}\binits{M.}},
  \bauthor{\bsnm{Nagata},~\bfnm{T.}\binits{T.}},
  \bauthor{\bsnm{Ikeda},~\bfnm{H.}\binits{H.}},
  \bauthor{\bsnm{Kunihiro},~\bfnm{N.}\binits{N.}},
  \bauthor{\bsnm{Takami},~\bfnm{T.}\binits{T.}} \AND
  \bauthor{\bsnm{Ohata},~\bfnm{K.}\binits{K.}}
(\byear{2012}).
\btitle{s{L}{O}{R}{E}{T}{A}-qm for interictal {M}{E}{G} epileptic spike
  analysis: {C}omparison of location and quantity with equivalent dipole
  estimation}.
\bjournal{Clinical Neurphysiology}
\bvolume{123}
\bpages{1496-1501}.
\end{barticle}
\endbibitem

\end{thebibliography}

\end{document}